\documentclass[a4paper,fleqn,usenatbib]{mnras}

\usepackage{newtxtext,newtxmath}
\usepackage[T1]{fontenc}
\usepackage{ae,aecompl}
\usepackage{bm}

\usepackage{graphicx}	% Including figure files
\usepackage{amsmath}	% Advanced maths commands
\usepackage{amssymb}	% Extra maths symbols

\usepackage{multicol}
\usepackage{verbatim}   % For commenting out multiple lines. --JR
\usepackage{threeparttable}
\usepackage{upgreek}

\newcommand{\lcdm}{$\Lambda$CDM}
\newcommand{\pcdm}{$\phi$CDM}
\newcommand{\om}{\Omega_{m0}}
\newcommand{\ol}{\Omega_{\Lambda}}
\newcommand{\ok}{\Omega_{k0}}
\newcommand{\FT}[1]{}

\title[Constraints from QSO, $H(z)$, and BAO data]{Quasar X-ray and UV flux, baryon acoustic oscillation, and Hubble parameter measurement constraints on cosmological model parameters}

\author[N. Khadka, B. Ratra]{
 Narayan Khadka,$^{1}$\thanks{E-mail: nkhadka@phys.ksu.edu}
and Bharat Ratra,$^{1}$\thanks{E-mail: ratra@phys.ksu.edu}
\\
$^{1}$Department of Physics, Kansas State University, 116 Cardwell Hall, Manhattan, KS 66502, USA\\
}

\date{Accepted XXX. Received YYY; in original form ZZZ}

\pubyear{2019}

\begin{document}
\label{firstpage}
\pagerange{\pageref{firstpage}--\pageref{lastpage}}
\maketitle

\begin{abstract}
We use the 2015 Risaliti and Lusso compilation of 808 X-ray and UV flux measurements of quasars (QSOs) in the redshift range $0.061 \leq z \leq 6.28$, alone and in conjunction with baryon acoustic oscillation (BAO) and Hubble parameter [$H(z)$] measurements, to constrain cosmological parameters in six cosmological models. The QSO data constraints are significantly weaker than, but consistent with, those from the $H(z)$ + BAO data. A joint analysis of the QSO + $H(z)$ + BAO data is consistent with the current standard model, spatially-flat $\Lambda$CDM, but mildly favors closed spatial hypersurfaces and dynamical dark energy.
%%%
\end{abstract}

\begin{keywords}
\textit{(cosmology:)} cosmological parameters -- \textit{(cosmology:)} observations -- \textit{(cosmology:)} dark energy
\end{keywords}

%%%%%%%%%%%%%%%%%%%%%%%%%%%%%%%%%%%%%%%%%%%%%%%%%%

%%%%%%%%%%%%%%%%% BODY OF PAPER %%%%%%%%%%%%%%%%%%

\section{Introduction}
\label{sec:Introduction}
Type Ia supernova (SNIa) apparent magnitude measurements provided the first convincing evidence for accelerated cosmological expansion (see \cite{Scolnic2018} for a recent discussion). Supporting evidence soon came from other cosmological probes, the most significant being cosmic microwave background (CMB) anisotropy data \citep{Plank2018}, baryon acoustic oscillation (BAO) distance measurements \citep{alam}, and Hubble parameter [$H(z)$] observations \citep{Moresco2016, Farooq2017}. If general relativity is an accurate model of gravitation, hypothetical dark energy is responsible for the observed acceleration of the cosmological expansion. There are many different dark energy models. In this paper we consider three of them and also consider flat and non-flat spatial hypersurfaces in each case, for a total of six cosmological models.

The simplest observationally-consistent dark energy model is the flat $\Lambda$CDM model, the current standard model  (Peebles 1984). In this model the accelerated expansion is powered by the spatially homogenous cosmological constant ($\Lambda$) energy density which is constant in time. This model is consistent with most observations when about $70\%$ of the current cosmological energy budget is contributed by dark energy, with about 25$\%$ coming from the cold dark matter (CDM), and the remaining 5$\%$ due to baryons. The standard model assumes flat spatial hypersurfaces. Current observations allow a little spatial curvature,\footnote{For discussion of observational constraints on spatial curvature, see \cite{Farooq2015}, \cite{chen6}, \cite{Yu2016}, \cite{wei2017}, \cite{Rana2017}, \cite{Ooba2018a, Ooba2018b, Ooba2018c}, \cite{DESa}, \cite{Witzemann2018}, \cite{Yu2018}, \cite{Park2018a, Park2018b, Park2018c, Park2018d, Park2019}, \cite{Mitra2019a}, \cite{penton2018}, \cite{Xu2019}, \cite{Zheng2019}, \cite{Ruan2019}, \cite{Giambo2019}, \cite{Cole2019}, \cite{Eingorn2019}, \cite{Jesus2019}, \cite{Handley2019}, and references therein.} so we can generalise the standard model to the non-flat $\Lambda$CDM model which allows for non-zero spatial curvature energy density.

While the $\Lambda$CDM model is consistent with many observations, its assumption of a time-independent and spatially-homogeneous dark energy density is difficult to theoretically motivate. Also, observations do not require that the dark energy density be time independent, and models in which the dark energy density decreases with time have been studied. Here we consider two dynamical dark energy models, the XCDM parametrization in which an $X$-fluid is the dynamical dark energy and the $\phi$CDM model in which a scalar field $\phi$ is the dynamical dark energy. We also study spatially flat and non-flat versions of both the XCDM parametrization and the $\phi$CDM model.

The main goal of our paper is to use the \cite{Risaliti2015} quasar (QSO) X-ray and UV flux measurements to constrain cosmological parameters. \cite{Risaliti2015} consider cosmological parameter constraints in the non-flat $\Lambda$CDM model; here we also consider cosmological parameter constraints in five other cosmological models. In addition, we examine the effect of different Hubble constant priors on the cosmological parameter constraints. By studying constraints in a number of models, we are able to draw somewhat model-independent conclusions about the QSO data constraints. We find that the \cite{Risaliti2015} QSO data by themselves do not provide very restrictive constraints on cosmological parameters. However, the QSO constraints are largely consistent with those that follow from the $H(z)$ + BAO data, and when jointly analyzed the QSO data slightly tighten and shift the $H(z)$ + BAO data constraints.

The QSO + $H(z)$ + BAO data are consistent with the standard flat $\Lambda$CDM cosmological model although they mildly favor closed spatial hypersurfaces over flat ones and dynamical dark energy over a cosmological constant.

While current QSO data by themselves do not provide restrictive cosmological parameter constraints, the new \cite{Risaliti2019} compilation of 1598 QSO measurements will provide tighter constraints, that should be improved upon by near-future QSO data. Currently, CMB anisotropy, BAO, SNIa, and $H(z)$ data provide the most restrictive constraints on cosmological parameters. To test consistency, and to help tighten cosmological parameter constraints, it is essential that additional cosmological probes, such as the QSO data studied here, be developed. 

This paper is organized as follows. In Sec. 2 we describe the models that we use. In Sec. 3 we discuss the data that we use to constrain cosmological parameters in these models. In Sec. 4 we describe the methodology adopted for these analyses. In Sec. 5 we present our results and conclude in Sec. 6.

%%%
\section{Models}
\label{sec:models}
In this paper we constrain cosmological parameters of the spatially-flat and non-flat versions of three different dark energy cosmological models, for a total of six cosmological models. For the dark energy we consider a cosmological constant $\Lambda$ in the $\Lambda$ cold dark matter ($\Lambda$CDM) model, as well as a decreasing dark energy density modeled as an $X$-fluid in the XCDM parametrization, or as a scalar field $\phi$ in the $\phi$CDM model.   

In the $\Lambda$CDM model the Hubble parameter, as a function of redshift $z$, is
\begin{equation}
\label{eq:friedLCDM}
    H(z) = H_0\sqrt{\Omega_{m0}(1 + z)^3 + \Omega_{k0}(1 + z)^2 + \Omega_{\Lambda}},
\end{equation}
and
\begin{equation}
\label{eq:friedLCDM}
    \om+\ok+\ol = 1.
\end{equation}
Here $H_0$ is the Hubble constant, $\om$ and $\ok$ are the current values of the non-relativistic matter and the spatial curvature energy density parameters and $\ol$ is the dark energy density parameter. For the spatially-flat $\Lambda$CDM model the free parameters are chosen to be $\om$ and $H_{0}$. For the spatially non-flat $\Lambda$CDM model the free parameters are chosen to be  $\om$, $\ol$, and $H_0$.

In the XCDM parametrization, the dark energy density is dynamical and decreases with time. The equation of state for the dark energy fluid is $P_X$ = $\omega_{X}$ $\rho_{X}$. Here $P_X$ is the pressure of the $X$-fluid, $\rho_{X}$ is the energy density of that fluid, and $\omega_{X}$ is the equation of state parameter whose value is negative ($\omega_X < -1/3$). In this model the Hubble parameter is
\begin{equation}
\label{eq:XCDM}
    H(z) = H_0\sqrt{\Omega_{m0}(1 + z)^3 + \Omega_{k0}(1 + z)^2 + \Omega_{X0}(1+z)^{3(1+\omega_X)}},
\end{equation}
and
\begin{equation}
\label{eq:XCDM}
    \om+\ok+\Omega_{X0} = 1.
\end{equation}
Here $\Omega_{X0}$ is the current value of the $X$-fluid energy density parameter. For the spatially-flat case the free parameters are $\om$, $\omega_X$, and $H_0$. For the non-flat case the free parameters are $\om$, $\ok$, $\omega_X$, and $H_0$. In the $\omega_{X}$ = $-1$ limit the XCDM model is the $\Lambda$CDM model.

In the $\phi$CDM model dark energy is modeled as a scalar field $\phi$ with potential energy density $V(\phi)$ \citep{peebles1988, Ratra1988, Pavlov2013}.\footnote{Discussion of constraints on the $\phi$CDM model may be traced through  \cite{chen2}, \cite{Samushia2007}, \cite{yashar2009}, \cite{SamushiaR2010}, \cite{Samushia2010}, \cite{chen4}, \cite{camp}, \cite{Farooq2013b}, \cite{Farooq2013a}, \cite{Avsa}, \cite{Sola2017}, \cite{Sola2018, Sola2019}, \cite{Zhai2017}, \cite{Sangwan2018}, \cite{Singh2019}, \cite{Mitra2019b}, and references therein.} In this model the dark energy density is dynamical and decreases with time. A widely used $V(\phi)$ is of the inverse power law form,
\begin{equation}
\label{eq:phiCDMV}
    V(\phi) = \frac{1}{2}\kappa m_{p}^2 \phi^{-\alpha},
\end{equation}
where $m_{p}$ is the Planck mass, $\alpha$ is a positive parameter, and 
\begin{equation}
\label{eq:kappa}
  \kappa = \frac{8}{3}\left(\frac{\alpha + 4}{\alpha + 2}\right)\left[\frac{2}{3}\alpha(\alpha + 2)\right]^{\alpha/2} .
\end{equation}
In this model the equations of motion are 
\begin{equation}
\label{field}
    \ddot{\phi} + \frac{3\dot{a}}{a}\dot\phi - \frac{1}{2}\alpha \kappa m_{p}^2 \phi^{-\alpha - 1} = 0,
\end{equation}
and,
\begin{equation}
\label{friedpCDM}
    \left(\frac{\dot{a}}{a}\right)^2 = \frac{8 \uppi G}{3}\left(\rho_m + \rho_{\phi}\right) - \frac{k}{a^2},
\end{equation}
where $a$ is the scale factor, overdots denote derivatives with respect to time, $k$ is positive, zero, and negative for closed, flat, and open spatial hypersurfaces, $\rho_m$ is the non-relativistic matter density, and the scalar field energy density is
\begin{equation}
    \rho_{\phi} = \frac{m^2_p}{32\pi}[\dot{\phi}^2 + \kappa m^2_p \phi^{-\alpha}].
\end{equation}
So, the Hubble parameter in this model is
\begin{equation}
    H(z) = H_0\sqrt{\Omega_{m0}\left(1 + z\right)^3 + \Omega_{k0}\left(1 + z\right)^2 + \Omega_{\phi}\left(z, \alpha\right)},
\end{equation}
where the scalar field energy density parameter 
\begin{equation}
    \Omega_{\phi}(z, \alpha) = \frac{8 \uppi G \rho_{\phi}}{3H^2_0},
\end{equation}
where $G$ is the gravitational constant, and $\Omega_{\phi}(z,\alpha)$  has to be numerically computed. For the spatially non-flat $\phi$CDM  model the free parameters are $\om$, $\ok$, $\alpha$, and $H_0$. For the spatially-flat $\phi$CDM model the free parameters are $\om$, $\alpha$, and $H_0$. In the limit $\alpha\rightarrow0$, the $\phi$CDM model reduces to the $\Lambda$CDM model.
%%%
%Old section
%%%
%%%
%end old section
%%%

\section{Data}
\label{sec:data}
%%%
We use three different data sets to constrain cosmological parameters. The main purpose of our paper is to use the 808 QSO X-ray and UV flux measurements of \cite{Risaliti2015},\footnote{Also see \cite{Risaliti2016}, \cite{Lusso2017}, and \cite{Biso}. For a newer compilation of QSO data see \cite{Risaliti2018}. For cosmological parameter constraints derived from QSO data, also see \cite{Lopez2016}, \cite{Lusso2019}, \cite{Melia2019}, and \cite{Lazkoz2019}.} extending over a redshift range of $0.061 \leq z \leq 6.28$, to determine cosmological parameter constraints, and to compare these QSO cosmological parameter constraints to those determined from more widely used BAO distance measurements and $H(z)$ observations. The BAO and $H(z)$ data we use are listed in Tables 1 and 2 of Ryan et al. (2018) and consist of 11 BAO measurements over the redshift range $0.106 \leq z \leq 2.36$ and 31 $H(z)$ measurements over the redshift range $0.07 \leq z \leq 1.965$.

\section{Method}
\label{sec:methods}
As described in \cite{Risaliti2015}, the method of analysis depends on the non-linear relation between the X-ray and UV luminosities of quasars. This relation is
\begin{equation}
\label{eq:XCDM}
    \log(L_{X}) = \beta + \gamma \log(L_{UV}) ,
\end{equation}
where $\log$ = $\log_{10}$  and $L_X$ and $L_{UV}$ are the QSO X-ray and UV luminosities. $\beta$ and $\gamma$ are  free parameters to be determined by using the data. 

Expressing the luminosity in terms of the flux, we obtain
\begin{equation}
\label{eq:XCDM}
    \log(F_{X}) = \beta +(\gamma - 1)\log(4\pi) + \gamma \log(F_{UV}) + 2(\gamma - 1)\log(D_L),
\end{equation}
where $F_X$ and $F_{UV}$ are the X-ray and UV fluxes respectively. Here $D_L$ is the luminosity distance, which is a function of redshift and cosmological parameters, which will allow us to constrain the cosmological model parameters. The luminosity distance $D_L(z, p)$ is given by
\begin{equation}
\label{eq:DM}
  \frac{H_0\sqrt{\left|\Omega_{k0}\right|}D_L(z, p)}{(1+z)} = 
    \begin{cases}
    {\rm sinh}\left[g(z)\right] & \text{if}\ \Omega_{k0} > 0, \\
    \vspace{1mm}
    g(z) & \text{if}\ \Omega_{k0} = 0,\\
    \vspace{1mm}
    {\rm sin}\left[g(z)\right] & \text{if}\ \Omega_{k0} < 0,
    \end{cases}   
\end{equation}
where $p$ is the set of cosmological model parameters,
\begin{equation}
\label{eq:XCDM}
   g(z) = H_0\sqrt{\left|\Omega_{k0}\right|}\int^z_0 \frac{dz'}{H(z')},
\end{equation}
and $H(z)$, which is a function of cosmological model parameters, is given in Sec. 2 for the six cosmological models we study in this paper.

We determine the best-fit values and uncertainty of the parameters for a given model by maximizing the likelihood function. For QSO data, we have the observed X-ray flux and we can predict the X-ray flux at given redshift as a function of cosmological parameters by using eqs. (13) and (14). So, the likelihood function $({\rm LF})$ for QSO data is
\begin{equation}
\label{eq:chi2}
    \ln({\rm LF}) = -\frac{1}{2}\sum^{808}_{i = 1} \left[\frac{[\log(F^{\rm obs}_{X,i}) - \log(F^{\rm th}_{X,i})]^2}{s^2_i} + \ln(2\pi s^2_i)\right],
\end{equation}
where $\ln$ = $\log_e$ and $s^2_i = \sigma^2_i + \delta^2$, where $\sigma_i$ and $\delta$ are the measurement error on $F^{\rm obs}_{X,i}$ and the global intrinsic dispersion respectively. We treat $\delta$ as a free parameter to be determined by the data, along with the other two free parameters, $\beta$ and $\gamma$, which characterise the $L_X$ - $L_{UV}$ relation in eq. (12). In eq. (16) $F^{\rm th}_{X,i}$ is the corresponding model prediction defined through eq. (13), and is a function of $F_{UV}$ and $D_L(z_i, p)$.

Our determination of the BAO and $H(z)$ data constraints follows \cite{Ryan2019}. The likelihood function for the uncorrelated BAO and $H(z)$ data is
\begin{equation}
\label{eq:chi2}
    \ln({\rm LF}) = -\frac{1}{2}\sum^{N}_{i = 1} \frac{[A_{\rm obs}(z_i) - A_{\rm th}(z_i, p)]^2}{\sigma^2_i},
\end{equation}
where $A_{\rm obs}(z_i)$ and $\sigma_i$ are the measured quantity and error bar at redshift $z_i$ and $A_{\rm th}(z_i, p)$ is the corresponding model-predicted value. The measurements in the first six lines of Table 1 of \cite{Ryan2019} are correlated and the likelihood function for those data points is
\begin{equation}
\label{eq:chi2}
    \ln({\rm LF}) = -\frac{1}{2} [A_{\rm obs}(z_i) - A_{\rm th}(z_i, p)]^T C^{-1} [A_{\rm obs}(z_i) - A_{\rm th}(z_i, p)],
\end{equation}
where $C^{-1}$ is the inverse of the covariance matrix $C$ \citep{Ryan2019} =
\begin{equation}
\label{covmat}
\begin{bmatrix}
    624.707 & 23.729 & 325.332 & 8.34963 & 157.386 & 3.57778 \\
    23.729 & 5.60873 & 11.6429 & 2.33996 & 6.39263 & 0.968056 \\
    325.332 & 11.6429 & 905.777 & 29.3392 & 515.271 & 14.1013 \\
    8.34963 & 2.33996 & 29.3392 & 5.42327 & 16.1422 & 2.85334 \\
    157.386 & 6.39263 & 515.271 & 16.1422 & 1375.12 & 40.4327 \\
    3.57778 & 0.968056 & 14.1013 & 2.85334 & 40.4327 & 6.25936 \\
\end{bmatrix}.
\end{equation}

For all parameters except for $H_0$, we assume top hat priors, non-zero over $0 \leq \om \leq 1$, $0 \leq \ol \leq 1.3$, $-0.7 \leq k \leq 0.7$, $-5 \leq \omega_X \leq 5$, $0 \leq \alpha \leq 3$ ($0 \leq \alpha \leq 1.2$ for QSO only), $-20 \leq \ln{\delta} \leq 10$, $0 \leq \beta \leq 11$, and $-2 \leq \gamma \leq 2$. Here $k$ = $-\Omega_{k0} a^2_0$ where $a_0$ is the current value of the scale factor. For the Hubble constant we use two different Gaussian priors, $H_0 = 68 \pm 2.8$ km s$^{-1}$ Mpc$^{-1}$ corresponding to the results of a median statistics analysis of a large compilation of $H_0$ measurements \citep{chen3},\footnote{This is consistent with earlier median statistics analyses \citep{Gott2001, chen1}, as well as with many other recent measurement of $H_0$ \citep{L'Huillier2017, chen5, Wang2017, Lin2017, DESb, Yu2018, Gomez2018, Haridasu2018, Zhang2018a, Zhang2018b, D}.} and $H_0 = 73.24 \pm 1.74$ km s$^{-1}$ Mpc$^{-1}$ from a recent local expansion rate measurement \citep{Riess2016}.\footnote{Other local expansion rate observations find slightly lower $H_0$ values and have somewhat larger error bars \citep{Rigault2015, Zhang2017, Dhaw, Fernandez2018, Freedman2019}, but see \cite{Yuan2019}.}

The likelihood analysis is performed using the Markov chain Monte Carlo (MCMC) method as implemented in the emcee package \citep{Foreman2013} in Python 3.7. By using the maximum likelihood value $\rm LF_{\rm max}$ we compute the minimum $\chi^2_{\rm min}$ value $-2\ln{(\rm LF_{\rm max})}$. In addition to $\chi^2_{\rm min}$ we also use the Akaike Information Criterion
\begin{equation}
\label{eq:AIC}
    AIC = \chi^2_{\rm min} + 2d 
\end{equation}
and the Bayes Information Criterion
\begin{equation}
\label{eq:AIC}
    BIC = \chi^2_{\rm min} + d\ln{N} 
\end{equation}
\citep{Ryan2018}, where $d$ is the number of free parameters, and $N$ is the number of data points. The $AIC$ and $BIC$ penalize models with a larger number of free parameters.
\begin{table*}
	\centering
	\caption{Unmarginalized best-fit parameters of all models for the $H_0 = 68 \pm 2.8$ km s$^{-1}$ Mpc$^{-1}$ prior.}
	\label{tab:BFP}
	\begin{threeparttable}
	\begin{tabular}{lccccccccccccc} % four columns, alignment for each
		\hline
		Model & Data set & $\om$ & $\ol$ & $\ok$ & $\omega_{X}$ & $\alpha$ & $H_0$\tnote{a} & $\delta$ & $\beta$ & $\gamma$ & $\chi^2_{\rm min}$ & $AIC$ & $BIC$\\
		\hline
		Flat \lcdm\ &  $H(z)$ + BAO & 0.29 & 0.71 & - & - & - & 67.56 & - & - & - & 32.47 & 36.47 & 39.95\\
		 & QSO & 0.20 & 0.80 & - & - & - & 68.00 & 0.32 & 8.29 & 0.59 & 468.94 & 478.94 & 502.41\\
		 & QSO + $H(z)$ + BAO & 0.30 & 0.70 & - & - & - & 67.97 & 0.32 & 8.53 & 0.58 & 497.01 & 507.01 & 530.74\\
		\hline
		Non-flat \lcdm\ &  $H(z)$ + BAO & 0.30 & 0.70 & $0.00$ & - & - & 68.23 & - & - & - & 27.05 & 33.05 & 38.26\\
		 & QSO & 0.12 & 1.13 & $-0.25$ & - & - & 68.00 & 0.32 & 8.57 & 0.58 & 466.13 & 478.13 & 506.30\\
		 & QSO + $H(z)$ + BAO & 0.30 & 0.70 & $0.00$ & - & - & 68.33 & 0.32 & 8.52 & 0.58 & 496.52 & 508.52 & 536.99\\
		\hline
		Flat XCDM &  $H(z)$ + BAO & 0.30 & 0.70 & - & $-0.96$ & - & 67.24 & - & - & - & 27.29 & 33.29 & 38.50\\
		 & QSO & 0.21 & 0.79 & - & $-1.69$ & - & 68.00 & 0.32 & 8.41 & 0.59 & 468.35 & 480.35 & 508.52\\
		 & QSO + $H(z)$ + BAO & 0.30 & 0.70 & - & $-0.98$ & - & 67.62 & 0.32 & 8.53 & 0.58 & 496.90 & 508.90 & 537.37\\
		 \hline
		Non-flat XCDM & $H(z)$ + BAO & 0.32 & - & $-0.23$ & $-0.74$ & - & 67.42 & - & - & - & 24.91 & 32.91 & 39.86\\
		 & QSO & 0.021 & - & $-0.30$ & $-0.67$ & - & 68.00 & 0.32 & 8.65 & 0.58 & 463.10 & 477.10 & 509.96\\
		 & QSO + $H(z)$ + BAO & 0.32 & - & $-0.19$ & $-0.78$ & - & 67.76 & 0.32 & 8.76 & 0.57 & 494.65 & 508.65 & 541.87\\
		\hline
		Flat \pcdm\ & $H(z)$ + BAO & 0.32 & - & - & - & 0.10 & 67.23 & - & - & - & 27.42 & 33.42 & 38.63\\
		 & QSO & 0.2 & - & - & - & 0.07 & 68.00 & 0.32 & 8.31 & 0.59 & 469.04 & 481.04 & 509.21\\
		 & QSO + $H(z)$ + BAO & 0.30 & - & - & - & 0.03 & 66.69 & 0.32 & 8.86 & 0.57 & 497.03 & 509.03 & 537.50\\
		\hline
		Non-flat $\phi$CDM & $H(z)$ + BAO & 0.33 & - & $-0.20$ & - & 1.20 & 65.86 & - & - & - & 25.04 & 33.04 & 39.99\\
		 & QSO & 0.20 & - & $-0.01$ & - & 0.30 & 68.00 & 0.33 & 8.20 & 0.59 & 471.06 & 485.06 & 517.92\\
		 & QSO + $H(z)$ + BAO & 0.29 & - & $-0.18$ & - & 0.47 & 69.57 & 0.31 & 9.01 & 0.57 & 494.73 & 508.73 & 541.95\\
		 \hline
	\end{tabular}
	\begin{tablenotes}
    \item[a]${\rm km}\hspace{1mm}{\rm s}^{-1}{\rm Mpc}^{-1}$.
    \end{tablenotes}
    \end{threeparttable}
\end{table*}

\begin{table*}
	\centering
	\caption{Unmarginalized best-fit parameters of all models for the $H_0 = 73.24 \pm 1.74$ km s$^{-1}$ Mpc$^{-1}$ prior.}
	\label{tab:BFP}
	\begin{threeparttable}
	\begin{tabular}{lccccccccccccc} % four columns, alignment for each
		\hline
		Model & Data set & $\om$ & $\ol$ & $\ok$ & $\omega_{X}$ & $\alpha$ & $H_0$\tnote{a} & $\delta$ & $\beta$ & $\gamma$ & $\chi^2_{\rm min}$ & $AIC$ & $BIC$\\
		\hline
		Flat \lcdm\ &  $H(z)$ + BAO & 0.30 & 0.70 & - & - & - & 69.11 & - & - & - & 33.76 & 38.76 & 41.24\\
		 & QSO & 0.20 & 0.80 & - & - & - & 73.24 & 0.32 & 8.26 & 0.59 & 468.94 & 478.94 & 502.41\\
		 & QSO + $H(z)$ + BAO & 0.30 & 0.70 & - & - & - & 69.09 & 0.32 & 8.53 & 0.58 & 503.30 & 513.30 & 536.76\\
		\hline
		Non-flat \lcdm\ &  $H(z)$ + BAO & 0.30 & 0.78 & $-0.08$ & - & - & 71.56 & - & - & - & 28.80 & 34.80 & 40.01\\
		 & QSO & 0.12 & 1.13 & $-0.25$ & - & - & 73.24 & 0.32 & 8.55 & 0.58 & 466.13 & 478.13 & 506.30\\
		 & QSO + $H(z)$ + BAO & 0.30 & 0.78 & $-0.08$ & - & - & 71.66 & 0.32 & 8.61 & 0.58 & 497.85 & 509.85 & 538.32\\
		\hline
		Flat XCDM &  $H(z)$ + BAO & 0.29 & 0.71 & - & $-1.14$ & - & 71.27 & - & - & - & 30.68 & 36.68 & 41.89\\
		 & QSO & 0.21 & 0.79 & - & $-1.69$ & - & 73.24 & 0.32 & 8.39 & 0.59 & 468.35 & 480.35 & 508.52\\
		 & QSO + $H(z)$ + BAO & 0.29 & 0.71 & - & $-1.14$ & - & 71.37 & 0.32 & 8.51 & 0.58 & 499.84 & 511.84 & 540.31\\
		 \hline
		Non-flat XCDM & $H(z)$ + BAO & 0.32 & - & $-0.21$ & $-0.85$ & - & 71.22 & - & - & - & 28.17 & 36.17 & 43.12\\
		 & QSO & 0.021 & - & $-0.30$ & $-0.67$ & - & 68.00 & 0.32 & 8.65 & 0.58 & 463.10 & 477.10 & 509.96\\
		 & QSO + $H(z)$ + BAO & 0.31 & - & $-0.17$ & $-0.91$ & - & 72.14 & 0.32 & 8.72 & 0.58 & 498.07 & 512.07 & 545.29\\
		\hline
		Flat \pcdm\ & $H(z)$ + BAO & 0.33 & - & - & - & 0.09 & 69.31 & - & - & - & 33.36 & 39.36 & 44.57\\
		 & QSO & 0.2 & - & - & - & 0.13 & 73.24 & 0.32 & 8.32 & 0.59 & 469.04 & 481.04 & 509.21\\
		 & QSO + $H(z)$ + BAO & 0.30 & - & - & - & 0.05 & 70.20 & 0.33 & 8.98 & 0.57 & 506.97 & 518.97 & 547.44\\
		\hline
		Non-flat $\phi$CDM & $H(z)$ + BAO & 0.32 & - & $-0.22$ & - & 1.14 & 69.23 & - & - & - & 27.62 & 35.62 & 42.57\\
		 & QSO & 0.20 & - & $-0.01$ & - & 0.30 & 71.00 & 0.33 & 8.20 & 0.59 & 473.45 & 487.45 & 520.31\\
		 & QSO + $H(z)$ + BAO & 0.32 & - & $-0.21$ & - & 1.17 & 73.51 & 0.32 & 9.99 & 0.53 & 497.58 & 511.58 & 544.80\\
		 \hline
	\end{tabular}
	\begin{tablenotes}
    \item[a]${\rm km}\hspace{1mm}{\rm s}^{-1}{\rm Mpc}^{-1}$.
    \end{tablenotes}
    \end{threeparttable}
\end{table*}
\begin{table*}
	\centering
	\caption{Marginalized one-dimensional best-fit parameters with 1$\sigma$ confidence intervals for all models using BAO and $H(z)$ data.}
	\label{tab:BFP}
	\begin{threeparttable}
	\begin{tabular}{lcccccccccc} % four columns, alignment for each
		\hline
		$H_0$\tnote{a}\hspace{3mm}prior & Model & $\om$ & $\ol$ & $\ok$ & $\omega_{X}$ & $\alpha$ & $H_0$\tnote{a}\\
		\hline
		$H_0 = 68 \pm 2.8$ & Flat \lcdm\ & $0.29^{+0.01}_{-0.01}$ & - & - & - & - & $67.58^{+0.85}_{-0.85}$ \\
		& Non-flat \lcdm\ & $0.30^{+0.01}_{-0.01}$ & $0.70^{+0.05}_{-0.06}$ & $0.00^{+0.06}_{-0.07}$ & - & - & $68.17^{+1.80}_{-1.79}$\\
		& Flat XCDM & $0.30^{+0.02}_{-0.02}$ & - & - & $-0.97^{+0.09}_{-0.09}$ & - & $67.39^{+1.87}_{-1.84}$\\
		& Non-flat XCDM & $0.32^{+0.02}_{-0.02}$ & - & $-0.18^{+0.17}_{-0.21}$ & $-0.77^{+0.11}_{-0.17}$ & - & $67.42^{+1.84}_{-1.80}$\\
		&Flat \pcdm\ & $0.31^{+0.01}_{-0.01}$ & - & - & - & $0.20^{+0.21}_{-0.13}$ & $66.57^{+1.31}_{-1.46}$\\
		& Non-flat $\phi$CDM & $0.31^{+0.01}_{-0.01}$ & - & $-0.20^{+0.13}_{-0.17}$ & - & $0.86^{+0.55}_{-0.49}$ & $67.69^{+1.75}_{-1.74}$\\
		\hline
		$H_0 = 73.24 \pm 1.74$ & Flat \lcdm\ & $0.31^{+0.01}_{-0.01}$ & - & - & - & - & $69.12^{+0.81}_{-0.80}$\\
		& Non-flat \lcdm\ & $0.30^{+0.01}_{-0.01}$ & $0.78^{+0.04}_{-0.04}$ & $-0.08^{+0.05}_{-0.05}$ & - & - & $71.51^{+1.41}_{-1.40}$\\
		& Flat XCDM & $0.29^{+0.02}_{-0.01}$ & - & - & $-1.14^{+0.08}_{-0.08}$ & - & $71.32^{+1.49}_{-1.48}$\\
		& Non-flat XCDM & $0.32^{+0.02}_{-0.02}$ & - & $-0.17^{+0.16}_{-0.19}$ & $-0.88^{+0.14}_{-0.21}$ & - & $71.23^{+1.46}_{-1.46}$\\
		&Flat \pcdm\ & $0.31^{+0.01}_{-0.01}$ & - & - & - & $0.07^{+0.09}_{-0.04}$ & $68.91^{+0.98}_{-1.00}$\\
		& Non-flat $\phi$CDM & $0.32^{+0.01}_{-0.01}$ & - & $-0.25^{+0.12}_{-0.16}$ & - & $0.68^{+0.53}_{-0.46}$ & $71.14^{+1.39}_{-1.38}$\\
		\hline
	\end{tabular}
	\begin{tablenotes}
    \item[a]${\rm km}\hspace{1mm}{\rm s}^{-1}{\rm Mpc}^{-1}$.
    \end{tablenotes}
    \end{threeparttable}
\end{table*}
\begin{table*}
	\centering
	\caption{Marginalized one-dimensional best-fit parameters with 1$\sigma$ confidence intervals for all models using QSO data.}
	\label{tab:BFP}
	\begin{threeparttable}
	\begin{tabular}{lccccccccccc} % four columns, alignment for each
		\hline
		$H_0$\tnote{a}\hspace{3mm}prior & Model & $\om$ & $\ol$ & $\ok$ & $\omega_{X}$ & $\alpha$ & $H_0$\tnote{a} & $\delta$ & $\beta$ & $\gamma$ \\
		\hline
		$H_0 = 68 \pm 2.8$ & Flat \lcdm\ & $0.26^{+0.17}_{-0.11}$ & - & - & - & - & $68^{+2.8}_{-2.8}$ & $0.32^{+0.008}_{-0.008}$ & $8.42^{+0.57}_{-0.58}$ & $0.59^{+0.02}_{-0.02}$\\
		& Non-flat \lcdm\ & $0.24^{+0.16}_{-0.10}$ & $0.93^{+0.18}_{-0.39}$ & $-0.17^{+0.49}_{-0.34}$ & - & - & $68^{+2.8}_{-2.8}$ & $0.32^{+0.008}_{-0.008}$ & $8.62^{+0.62}_{-0.62}$ & $0.58^{+0.02}_{-0.02}$\\
		& Flat XCDM & $0.25^{+0.16}_{-0.10}$ & - & - & $-2.49^{+1.26}_{-1.59}$ & - & $68^{+2.8}_{-2.8}$ & $0.32^{+0.008}_{-0.008}$ & $8.65^{+0.55}_{-0.57}$ & $0.58^{+0.02}_{-0.02}$\\
		& Non-flat XCDM & $0.29^{+0.26}_{-0.14}$ & - & $0.11^{+0.66}_{-0.31}$ & $-1.87^{+1.18}_{-2.05}$ & - & $68^{+2.8}_{-2.8}$ & $0.32^{+0.008}_{-0.008}$ & $8.52^{+0.64}_{-0.65}$ & $0.58^{+0.02}_{-0.02}$\\
		&Flat \pcdm\ & $0.26^{+0.18}_{-0.11}$ & - & - & - & $0.54^{+0.43}_{-0.38}$ & $68^{+2.8}_{-2.8}$ & $0.32^{+0.008}_{-0.008}$ & $8.42^{+0.57}_{-0.57}$ & $0.59^{+0.02}_{-0.02}$\\
		& Non-flat $\phi$CDM & $0.34^{+0.24}_{-0.16}$ & - & $-0.3^{+0.44}_{-0.61}$ & - & $0.55^{+0.43}_{-0.38}$ & $68^{+2.8}_{-2.8}$ & $0.32^{+0.008}_{-0.008}$ & $8.45^{+0.57}_{-0.58}$ & $0.59^{+0.02}_{-0.02}$\\
		\hline
		$H_0 = 73 \pm 1.74$ & Flat \lcdm\ & $0.26^{+0.17}_{-0.11}$ & - & - & - & - & $73.24^{+1.73}_{-1.73}$ & $0.32^{+0.008}_{-0.008}$ & $8.40^{+0.57}_{-0.57}$ & $0.59^{+0.02}_{-0.02}$\\
		& Non-flat \lcdm\ & $0.24^{+0.16}_{-0.10}$ & $0.93^{+0.18}_{-0.39}$ & $-0.17^{+0.49}_{-0.34}$ & - & - & $73.24^{+1.73}_{-1.73}$ & $0.32^{+0.008}_{-0.008}$ & $8.59^{+0.62}_{-0.62}$ & $0.58^{+0.02}_{-0.02}$\\
		& Flat XCDM & $0.25^{+0.16}_{-0.10}$ & - & - & $-2.48^{+1.26}_{-1.59}$ & - & $73.24^{+1.73}_{-1.73}$ & $0.32^{+0.008}_{-0.008}$ & $8.62^{+0.55}_{-0.56}$ & $0.58^{+0.02}_{-0.02}$\\
		& Non-flat XCDM & $0.29^{+0.25}_{-0.14}$ & - & $0.10^{+0.62}_{-0.32}$ & $-1.83^{+1.15}_{-2.02}$ & - & $73.24^{+1.74}_{-1.74}$ & $0.32^{+0.008}_{-0.008}$ & $8.50^{+0.65}_{-0.64}$ & $0.58^{+0.02}_{-0.02}$\\
		&Flat \pcdm\ & $0.24^{+0.19}_{-0.12}$ & - & - & - & $0.55^{+0.43}_{-0.38}$ & $73.23^{+1.73}_{-1.73}$ & $0.32^{+0.008}_{-0.008}$ & $8.40^{+0.57}_{-0.57}$ & $0.59^{+0.02}_{-0.02}$\\
		& Non-flat $\phi$CDM & $0.34^{+0.24}_{-0.17}$ & - & $-0.30^{+0.62}_{-0.44}$ & - & $0.55^{+0.43}_{-0.38}$ & $73.26^{+1.74}_{-1.73}$ & $0.32^{+0.008}_{-0.008}$ & $8.42^{+0.57}_{-0.58}$ & $0.59^{+0.02}_{-0.02}$\\
		\hline
	\end{tabular}
	\begin{tablenotes}
    \item[a]${\rm km}\hspace{1mm}{\rm s}^{-1}{\rm Mpc}^{-1}$.
    \end{tablenotes}
    \end{threeparttable}
\end{table*}

\begin{table*}
	\centering
	\caption{Marginalized one-dimensional best-fit parameters with 1$\sigma$ confidence intervals for all models using  QSO+$H(z)$+BAO data.}
	\label{tab:BFP}
	\begin{threeparttable}
	\begin{tabular}{lcccccccccc} % four columns, alignment for each
		\hline
		$H_0$\tnote{a}\hspace{3mm}prior & Model & $\om$ & $\ol$ & $\ok$ & $\omega_{X}$ & $\alpha$ & $H_0$\tnote{a} & $\delta$ & $\beta$ & $\gamma$\\
		\hline
		$H_0 = 68 \pm 2.8$ & Flat \lcdm\ & $0.30^{+0.01}_{-0.01}$ & $0.70^{+0.01}_{-0.01}$ & - & - & - & $67.99^{+0.85}_{-0.84}$ & $0.32^{+0.008}_{-0.008}$ & $8.53^{+0.49}_{-0.48}$ & $0.58^{+0.02}_{-0.02}$\\
		& Non-flat \lcdm\ & $0.30^{+0.01}_{-0.01}$ & $0.70^{+0.05}_{-0.05}$ & $0.00^{+0.06}_{-0.06}$ & - & - & $68.28^{+1.50}_{-1.51}$ & $0.32^{+0.008}_{-0.008}$ & $8.51^{+0.49}_{-0.48}$ & $0.58^{+0.02}_{-0.02}$\\
		& Flat XCDM & $0.30^{+0.02}_{-0.01}$ & - & - & $-0.98^{+0.08}_{-0.08}$ & - & $67.69^{+1.56}_{-1.53}$ & $0.32^{+0.008}_{-0.008}$ & $8.53^{+0.49}_{-0.48}$ & $0.58^{+0.02}_{-0.02}$\\
		& Non-flat XCDM & $0.31^{+0.02}_{-0.02}$ & - & $-0.15^{+0.15}_{-0.16}$ & $-0.80^{+0.11}_{-0.16}$ & - & $67.73^{+1.54}_{-1.52}$ & $0.32^{+0.008}_{-0.008}$ & $8.73^{+0.53}_{-0.52}$ & $0.58^{+0.02}_{-0.02}$\\
		&Flat \pcdm\ & $0.31^{+0.01}_{-0.01}$ & - & - & - & $0.16^{+0.17}_{-0.10}$ & $66.93^{+1.14}_{-1.22}$ & $0.32^{+0.008}_{-0.008}$ & $8.56^{+0.48}_{-0.48}$ & $0.58^{+0.02}_{-0.02}$\\
		& Non-flat $\phi$CDM & $0.31^{+0.01}_{-0.01}$ & - & $-0.18^{+0.11}_{-0.14}$ & - & $0.74^{+0.48}_{-0.43}$ & $67.87^{+1.49}_{-1.49}$ & $0.32^{+0.008}_{-0.008}$ & $8.77^{+0.52}_{-0.51}$ & $0.57^{+0.02}_{-0.02}$\\
		\hline
		$H_0 = 73.24 \pm 1.74$ & Flat \lcdm\ & $0.31^{+0.01}_{-0.01}$ & $0.69^{+0.01}_{-0.01}$ & - & - & - & $69.86^{+0.75}_{-0.75}$ & $0.32^{+0.008}_{-0.008}$ & $8.54^{+0.48}_{-0.48}$ & $0.58^{+0.02}_{-0.02}$\\
		& Non-flat \lcdm\ & $0.30^{+0.01}_{-0.01}$ & $0.80^{+0.04}_{-0.04}$ & $-0.10^{+0.05}_{-0.05}$ & - & - & $72.26^{+1.09}_{-1.09}$ & $0.32^{+0.008}_{-0.008}$ & $8.62^{+0.49}_{-0.48}$ & $0.58^{+0.02}_{-0.02}$\\
		& Flat XCDM & $0.29^{+0.01}_{-0.01}$ & - & - & $-1.18^{+0.07}_{-0.07}$ & - & $72.20^{+1.14}_{-1.13}$ & $0.32^{+0.008}_{-0.008}$ & $8.50^{+0.48}_{-0.48}$ & $0.58^{+0.02}_{-0.02}$\\
		& Non-flat XCDM & $0.31^{+0.02}_{-0.02}$ & - & $-0.14^{+0.13}_{-0.15}$ & $-0.94^{+0.14}_{-0.20}$ & - & $72.14^{+1.13}_{-1.12}$ & $0.32^{+0.008}_{-0.008}$ & $8.70^{+0.53}_{-0.52}$ & $0.58^{+0.02}_{-0.02}$\\
		&Flat \pcdm\ & $0.32^{+0.01}_{-0.01}$ & - & - & - & $0.05^{+0.06}_{-0.03}$ & $69.96^{+0.83}_{-0.85}$ & $0.32^{+0.008}_{-0.008}$ & $8.54^{+0.48}_{-0.48}$ & $0.58^{+0.02}_{-0.02}$\\
		& Non-flat $\phi$CDM & $0.31^{+0.01}_{-0.01}$ & - & $-0.22^{+0.09}_{-0.13}$ & - & $0.51^{+0.43}_{-0.33}$ & $72.02^{+1.07}_{-1.09}$ & $0.32^{+0.008}_{-0.008}$ & $8.82^{+0.52}_{-0.52}$ & $0.57^{+0.02}_{-0.02}$\\
		\hline
	\end{tabular}
	\begin{tablenotes}
    \item[a]${\rm km}\hspace{1mm}{\rm s}^{-1}{\rm Mpc}^{-1}$.
    \end{tablenotes}
    \end{threeparttable}
\end{table*}
%%Note A

\section{Results}
\label{sec:Results}
\subsection{$H(z)$ + BAO constraints}
\label{sec:Hz+BAO}
Results for the $H(z)$ + BAO data set are listed in Tables 1--3. The unmarginalized best-fit parameters are in Tables 1 and 2. The marginalized one-dimensional best-fit parameter values with $1\sigma$ error bars are given in Table 3. These results are consistent with those of \cite{Ryan2019}. The slight differences between the two sets of results are the consequence of the different analysis techniques used and the Gaussian priors used for the Hubble constant. Our computations are done using the MCMC method while \cite{Ryan2019} used a grid-based $\chi^2$ technique.

The one-dimensional likelihoods and two-dimensional $1\sigma$, $2\sigma$, and $3\sigma$ confidence contours for all parameters determined by using $H(z)$ + BAO data are shown in red in Figs. 1--12. Some of the plots for the $\phi$CDM model differ slightly from the corresponding plots of \cite{Ryan2019} because of the difference we discussed above. The $H(z)$ + BAO data reduced $\chi^2$ values are $\sim$ 0.6--0.8, smaller than unity largely because of the $H(z)$ data.

\subsection{QSO constraints}
\label{sec:QSO}
Use of the QSO data to constrain cosmological parameters is based on the assumed validity  of the $L_X$ - $L_{UV}$ relation in eq. (12). This assumption is tested by \cite{Risaliti2015}. By fitting this relation in cosmological models we have found, in agreement with \cite{Risaliti2015}, that the slope $\gamma$ $\sim$ $0.60 \pm 0.02$, the intercept $\beta$ is between 8 and 9, and the global intrinsic dispersion $\delta$ = $0.32 \pm 0.08$. The global intrinsic dispersion is large and so cosmological parameter determination done using these data is not as precise as that done by using, for example, the SNIa data. But the main advantage of using the quasar sample is that it covers a very large redshift range and eventually with more and better quality data it should result in tight constraints.

The QSO data determined cosmological parameter results are given in Tables 1, 2, and 4. The unmarginalized best-fit parameters are given in the Tables 1 and 2 for the $H_0 = 68 \pm 2.8$ ${\rm km}\hspace{1mm}{\rm s}^{-1}{\rm Mpc}^{-1}$ and $73.24 \pm 1.74$ ${\rm km}\hspace{1mm}{\rm s}^{-1}{\rm Mpc}^{-1}$ priors respectively. The one-dimensional likelihoods and the two-dimensional confidence contours are shown in grey in the left panels of Figs.\ 1--12. The cosmological parameter constraints are insensitive to the $H_0$ prior used. \cite{Risaliti2015} have determined cosmological parameters for the non-flat $\Lambda$CDM model. Our QSO data constraints in the $\ol$ -- $\om$ sub-panels of our Figs. 3 and 4 agree well with the corresponding constraints in Fig. 6 of \cite{Risaliti2015}. For the non-flat $\Lambda$CDM model we find $\om$ = $0.24^{+0.16}_{-0.10}$ and $\ol$ = $0.93^{+0.18}_{-0.39}$, also in good agreement with the corresponding \cite{Risaliti2015} values of $\om$ = $0.22^{+0.10}_{-0.08}$ and $\ol$ = $0.92^{+0.18}_{-0.30}$.

The cosmological parameters obtained by using these QSO data have relatively high uncertainty for all models but they are mostly consistent with the results obtained by using the  BAO + $H(z)$ data set, which are shown in red in Figs.\ 1--12. The QSO data reduced $\chi^2$ values are also small $\sim$ 0.6--0.7; it is of interest to understand why this is so. The QSO data $\chi^2_{\rm min}$ values do not change significantly from model to model.

\subsection{QSO + $H(z)$ + BAO constraints}
\label{QSOBAOH2}
 From Figs.\ 1--12 we see that the constraints from $H(z)$ + BAO data and those from QSO data alone are largely mutually consistent, except for the $H_0 = 73.24 \pm 1.74$ ${\rm km}\hspace{1mm}{\rm s}^{-1}{\rm Mpc}^{-1}$ prior case in the flat $\Lambda$CDM and flat $\phi$CDM models, see the bottom left sub-panels in the left panels of Figs.\ 2 and 10. The $H(z)$ + BAO data constrain cosmological parameters quite tightly while the QSO data result in very loose constraints on these parameters. Although the QSO data alone are not able to provide restrictive constraints, they can help tighten constraints when used in combination with $H(z)$ + BAO data.
 
 Given that the QSO and the $H(z)$ + BAO constraints are mostly consistent, it is reasonable to determine joint QSO + $H(z)$ + BAO constraints. These results are given in Tables 1, 2, and 5. The QSO + $H(z)$ + BAO one-dimensional likelihoods and two-dimensional confidence contours for all the cosmological parameters are shown in blue in the right panels of Figs.\ 1--12. These figures also show the $H(z)$ + BAO data constraint contours in red. Adding the QSO data to the $H(z)$ + BAO data and deriving joint constraints on cosmological parameters, results in bigger effects for the case of the $H_0 = 73.24 \pm 1.74$ ${\rm km}\hspace{1mm}{\rm s}^{-1}{\rm Mpc}^{-1}$ prior (Figs. 2, 4, 6, 8, 10, $\&$ 12), and in the flat $\phi$CDM model for both priors (Figs.\ 9 $\&$ 10).

Using the combined data set we see, from Table 5, the non-relativistic matter density parameter is measured to lie in the range $\om$ = $0.30 \pm 0.01$ to $0.31 \pm 0.01$ ($\om$ = $0.30 \pm 0.01$ to $0.31 \pm 0.02$)  for flat (non-flat) models and the $H_0 = 68 \pm 2.8$ ${\rm km}\hspace{1mm}{\rm s}^{-1}{\rm Mpc}^{-1}$ prior and to lie in the range $\om$ = $0.29 \pm 0.01$ to $0.32 \pm 0.01$ ($\om$ = $0.30 \pm 0.01$ to $0.31 \pm 0.02$)  for flat (non-flat) models and the $H_0 = 73.24 \pm 1.74$ ${\rm km}\hspace{1mm}{\rm s}^{-1}{\rm Mpc}^{-1}$ prior. In some cases these results differ slightly from the $H(z)$ + BAO data results of Table 3. These results are consistent with those derived using other data.

The Hubble constant is found to lie in the range $H_0$ = $66.93^{+1.14}_{-1.22}$ to $67.99^{+0.85}_{-0.84}$ ($H_0$ = $67.73^{+1.54}_{-1.52}$ to $68.28^{+1.50}_{-1.51}$) ${\rm km}\hspace{1mm}{\rm s}^{-1}{\rm Mpc}^{-1}$ for flat (non-flat) models and the $H_0 = 68 \pm 2.8$ ${\rm km}\hspace{1mm}{\rm s}^{-1}{\rm Mpc}^{-1}$ prior and to lie in the range $H_0$ = $69.86 \pm 0.75$ to $72.20^{+1.14}_{-1.13}$ ($H_0$ = $72.02^{+1.07}_{-1.09}$ to $72.26 \pm 1.09$) ${\rm km}\hspace{1mm}{\rm s}^{-1}{\rm Mpc}^{-1}$ for flat (non-flat) models and the $H_0 = 73.24 \pm 1.74$ ${\rm km}\hspace{1mm}{\rm s}^{-1}{\rm Mpc}^{-1}$ prior. As expected, for the $H_0 = 73.24 \pm 1.74$ ${\rm km}\hspace{1mm}{\rm s}^{-1}{\rm Mpc}^{-1}$ prior case, the measured value of $H_0$ is pulled lower than the prior value because the $H(z)$ and BAO data favor a lower $H_0$.

For the non-flat $\Lambda$CDM model the curvature energy density parameter is measured to be $\ok$ = $0.00 \pm 0.06$ and $-0.10 \pm 0.05$ for the $H_0 = 68 \pm 2.8$ ${\rm km}\hspace{1mm}{\rm s}^{-1}{\rm Mpc}^{-1}$ and $73.24 \pm 1.74$ ${\rm km}\hspace{1mm}{\rm s}^{-1}{\rm Mpc}^{-1}$ priors respectively.  The curvature energy density parameter is found to be $\ok$ = $-0.15^{+0.15}_{-0.16}$ and $-0.18^{+0.11}_{-0.14}$ for the non-flat XCDM and non-flat $\phi$CDM models for the $H_0 = 68 \pm 2.8$ ${\rm km}\hspace{1mm}{\rm s}^{-1}{\rm Mpc}^{-1}$ prior and $\ok$ = $-0.14^{+0.13}_{-0.15}$ and $-0.22^{+0.09}_{-0.13}$ for the non-flat XCDM and non-flat $\phi$CDM models for the $H_0 = 73.24 \pm 1.74$ ${\rm km}\hspace{1mm}{\rm s}^{-1}{\rm Mpc}^{-1}$ prior. It is interesting that in all cases closed spatial hypersurfaces are favored, albeit just barely in the non-flat $\Lambda$CDM model with the $H_0 = 68 \pm 2.8$ ${\rm km}\hspace{1mm}{\rm s}^{-1}{\rm Mpc}^{-1}$ prior, and only at 1$\sigma$ for most other cases, but at more than 2$\sigma$ for the non-flat $\phi$CDM model and the $H_0 = 73.24 \pm 1.74$ ${\rm km}\hspace{1mm}{\rm s}^{-1}{\rm Mpc}^{-1}$ prior. This preference for closed spatial hypersurfaces is largely driven by the $H(z)$ + BAO data \citep{Park2018d, Ryan2019}. Mildly closed spatial hypersurfaces are also consistent with CMB anisotropy measurements \citep{Ooba2018a, Ooba2018b, Ooba2018c, Park2018a, Park2018b, Park2018c, Park2019}.

The cosmological constant density parameter for the flat (non-flat) $\Lambda$CDM model is determined to be $\ol$ = $0.70 \pm 0.01$($0.70 \pm 0.05$) and $0.69 \pm 0.01$ ($0.80 \pm 0.04$) for the $H_0 = 68 \pm 2.8$ ${\rm km}\hspace{1mm}{\rm s}^{-1}{\rm Mpc}^{-1}$ and $73.24 \pm 1.74$ ${\rm km}\hspace{1mm}{\rm s}^{-1}{\rm Mpc}^{-1}$ priors respectively.

The parameters that govern dark energy dynamics move closer to those of a time-independent $\Lambda$ when we jointly analyze the QSO data with the $H(z)$ + BAO data, compared to the corresponding QSO data alone case. From the analyses of the QSO + $H(z)$ + BAO data the equation of state parameter for the flat (non-flat) XCDM parametrization is determined to be $\omega_X$ = $-0.98 \pm 0.08$ ($-0.80^{+0.11}_{-0.16}$) and $-1.18 \pm 0.07$ ($-0.94^{+0.14}_{-0.20}$) for the $H_0 = 68 \pm 2.8$ ${\rm km}\hspace{1mm}{\rm s}^{-1}{\rm Mpc}^{-1}$ and $73.24 \pm 1.74$ ${\rm km}\hspace{1mm}{\rm s}^{-1}{\rm Mpc}^{-1}$ priors respectively. The parameter $\alpha$ in the flat (non-flat) $\phi$CDM model is determined to be $\alpha$ = $0.16^{+0.17}_{-0.10}$ ($0.74^{+0.48}_{-0.43}$) and $0.05^{+0.06}_{-0.03}$ ($0.51^{+0.43}_{-0.33}$) for the $H_0 = 68 \pm 2.8$ ${\rm km}\hspace{1mm}{\rm s}^{-1}{\rm Mpc}^{-1}$ and $73.24 \pm 1.74$ ${\rm km}\hspace{1mm}{\rm s}^{-1}{\rm Mpc}^{-1}$ priors respectively. Of these 8 cases, 6 favor dark energy dynamics over a time-independent cosmological constant energy density; for the $68 \pm 2.8$ ${\rm km}\hspace{1mm}{\rm s}^{-1}{\rm Mpc}^{-1}$ prior dynamical dark energy is favored at 1.3$\sigma$ to 1.7$\sigma$ in the flat $\phi$CDM and non-flat XCDM and $\phi$CDM models, while for the $73.24 \pm 1.74$ ${\rm km}\hspace{1mm}{\rm s}^{-1}{\rm Mpc}^{-1}$ prior, dark energy dynamics is favored at 1.5$\sigma$ to 2.6$\sigma$ in the flat and non-flat $\phi$CDM models and the flat XCDM cases. Other data also favor mild dark energy dynamics \citep{Ooba2018d, Park2018b, Park2018c}.

\section{Conclusion}
\label{con}
We have used the \cite{Risaliti2015} compilation of 808 X-ray and UV QSO flux measurements to constrain cosmological parameters in six cosmological models. These QSO data constraints are much less restrictive than, but mostly consistent with those obtained from the joint analyses of 31 Hubble parameter and 11 BAO distance measurements.

We find that joint analyses of the QSO and $H(z)$ + BAO data tightens (and in some models, alters) constraints on cosmological parameters derived using just the $H(z)$ + BAO data. In general, the tightening effect is more significant in models with a larger number of free parameters. The joint QSO + $H(z)$ + BAO data constraints are consistent with the current standard flat $\Lambda$CDM model, although they weakly favor closed over flat spatial hypersurfaces and dynamical dark energy over a cosmological constant.

While cosmological parameter constraints from the QSO data we have used here are not that restrictive, the new \cite{Risaliti2019} QSO data compilation (of 1598 measurements over the redshift range $0.04 \leq z \leq 5.1$) will result in more restrictive cosmological parameter constraints that near-future QSO data should improve upon.
\section*{Acknowledgements}
We thank Elisabeta Lusso for her generous help, and we thank Lado Samushia, Chan-Gyung Park, and Joe Ryan for useful discussions. We are grateful to the Beocat Research Cluster at Kansas State University team, especially Dave Turner and Adam Tygart. This research was supported in part by DOE grant DE-SC0019038.



\begin{figure*}
\begin{multicols}{1}
    \includegraphics[width=\linewidth]{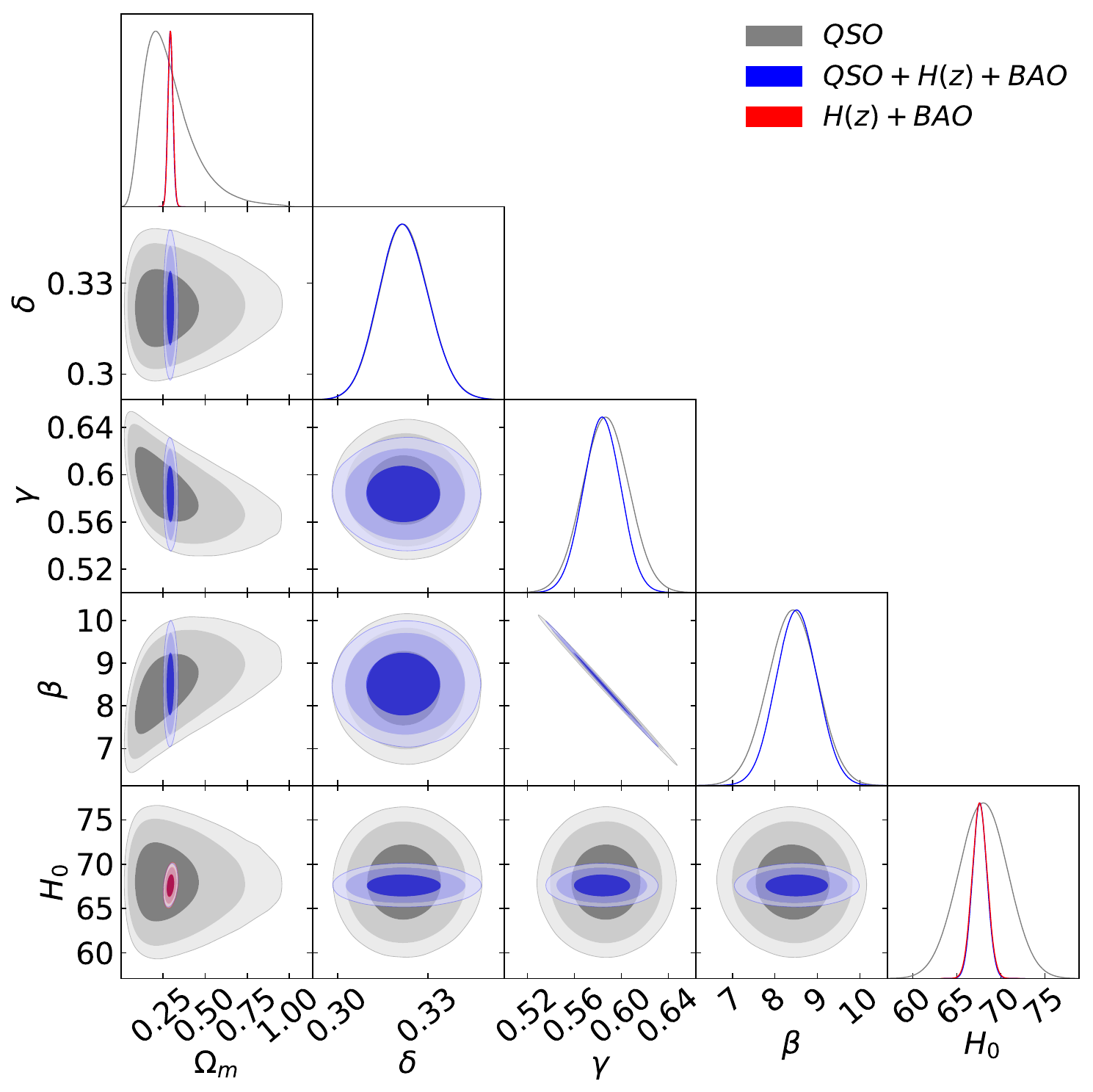}\par
    \includegraphics[width=\linewidth]{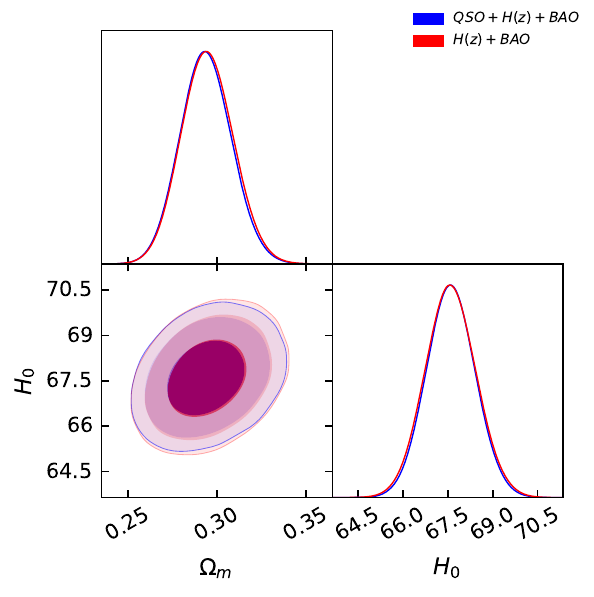}\par
\end{multicols}
\caption{Flat \lcdm\ model constraints from QSO (grey), $H(z)$ + BAO (red),  and QSO + $H(z)$ + BAO (blue) data. Left panel shows 1, 2, and 3$\sigma$ confidence contours and one-dimensional likelihoods for all free parameters. Right panel shows magnified plots for only cosmological parameters $\om$ and $H_0$, without the QSO-only constraints. These plots are for the $H_0 = 68 \pm 2.8$ ${\rm km}\hspace{1mm}{\rm s}^{-1}{\rm Mpc}^{-1}$ prior.}
\label{fig:flat LCDM68 model with BAO, H(z) and QSO data}
\end{figure*}
\begin{figure*}
\begin{multicols}{1}
    \includegraphics[width=\linewidth]{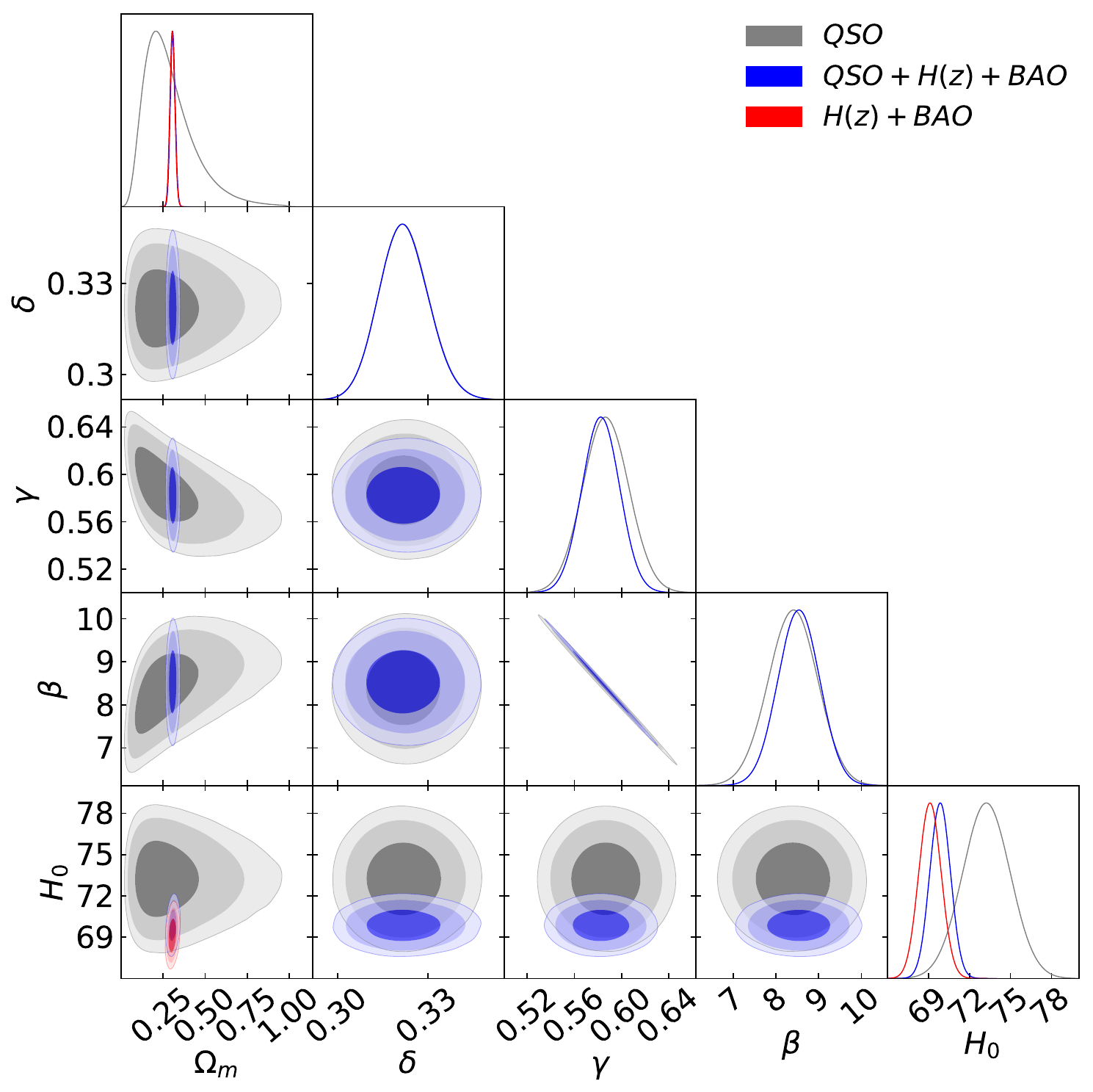}\par
    \includegraphics[width=\linewidth]{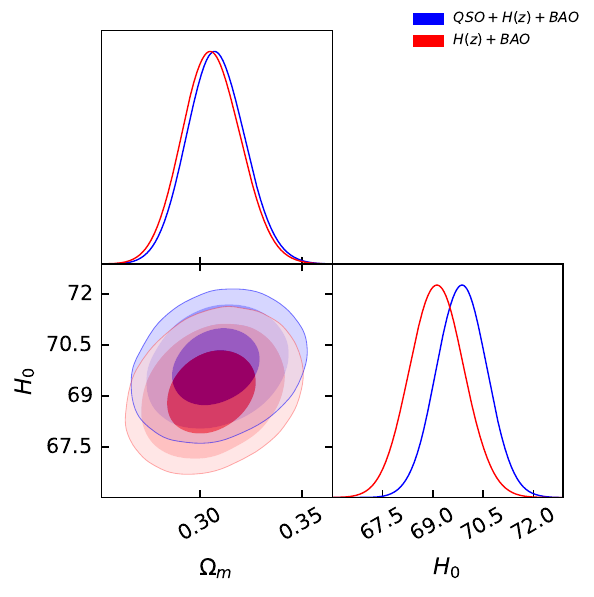}\par
\end{multicols}
\caption{Flat \lcdm\ model constraints from QSO (grey), $H(z)$ + BAO (red),  and QSO + $H(z)$ + BAO (blue) data. Left panel shows 1, 2, and 3$\sigma$ confidence contours and one-dimensional likelihoods for all free parameters. Right panel shows magnified plots for only cosmological parameters $\om$ and $H_0$, without the QSO-only constraints. These plots are for the $H_0 = 73.24 \pm 1.74$ ${\rm km}\hspace{1mm}{\rm s}^{-1}{\rm Mpc}^{-1}$ prior.}
\label{fig:flat LCDM73 model with BAO, H(z) and QSO data}
\end{figure*}
\begin{figure*}
\begin{multicols}{1}
    \includegraphics[width=\linewidth]{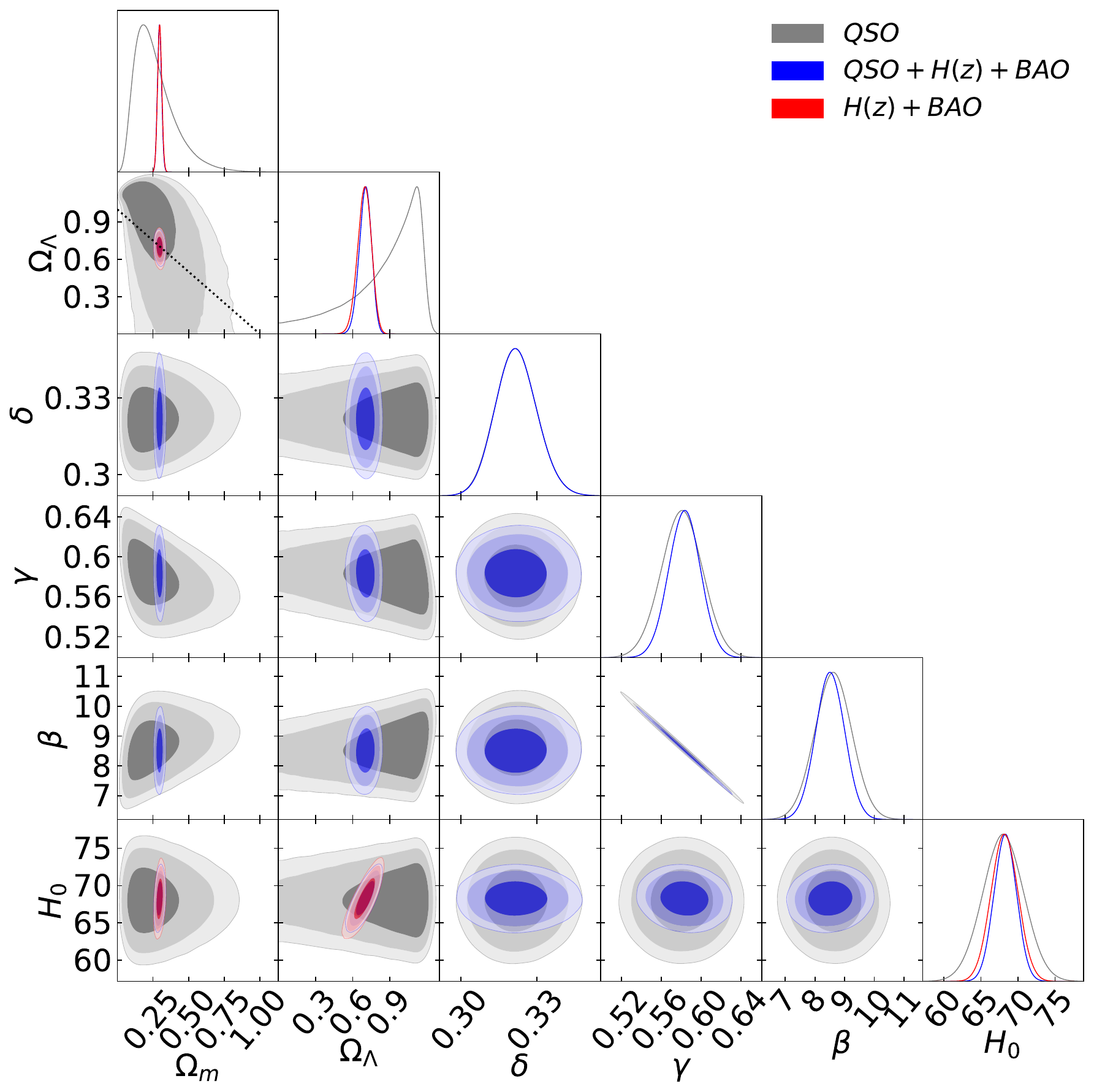}\par
    \includegraphics[width=\linewidth]{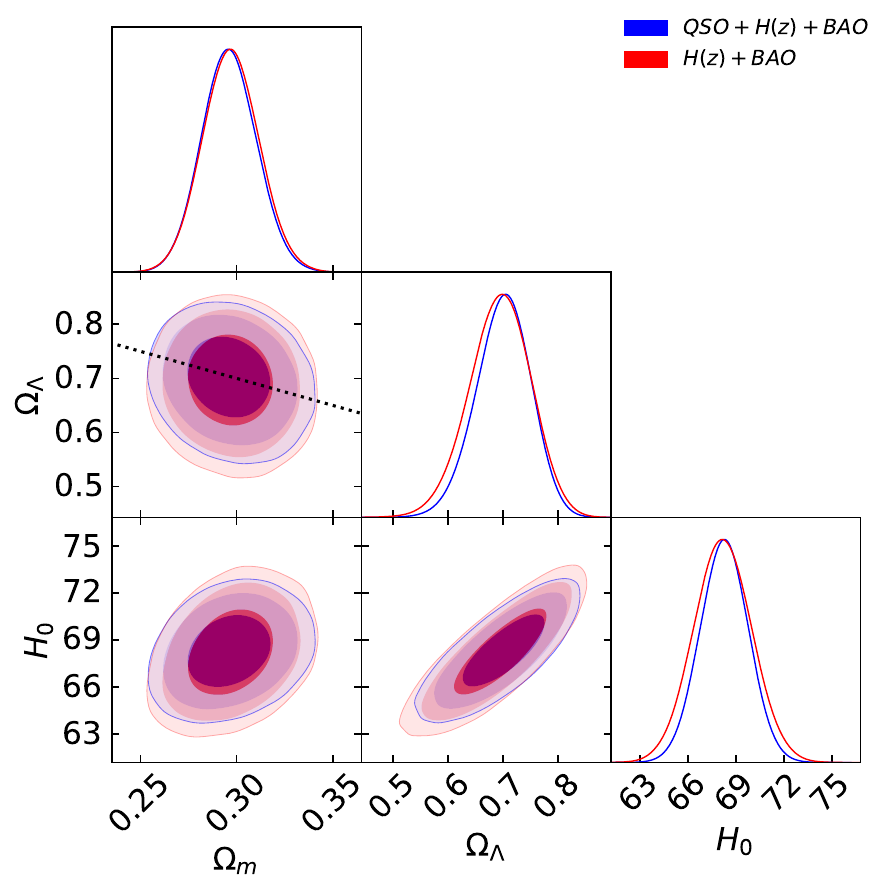}\par
\end{multicols}
\caption{Non-flat \lcdm\ model constraints from QSO (grey), $H(z)$ + BAO (red),  and QSO + $H(z)$ + BAO (blue) data. Left panel shows 1, 2, and 3$\sigma$ confidence contours and one-dimensional likelihoods for all free parameters. Right panel shows magnified plots for cosmological parameters $\om$, $\ol$, and $H_0$, without the QSO-only constraints. These plots are for the $H_0 = 68 \pm 2.8$ ${\rm km}\hspace{1mm}{\rm s}^{-1}{\rm Mpc}^{-1}$ prior. The black dotted straight lines corresponds to the flat $\Lambda$CDM model.}
\label{fig:non-flat LCDM68 model with BAO, H(z) and QSO data}
\end{figure*}
\begin{figure*}
\begin{multicols}{1}
    \includegraphics[width=\linewidth]{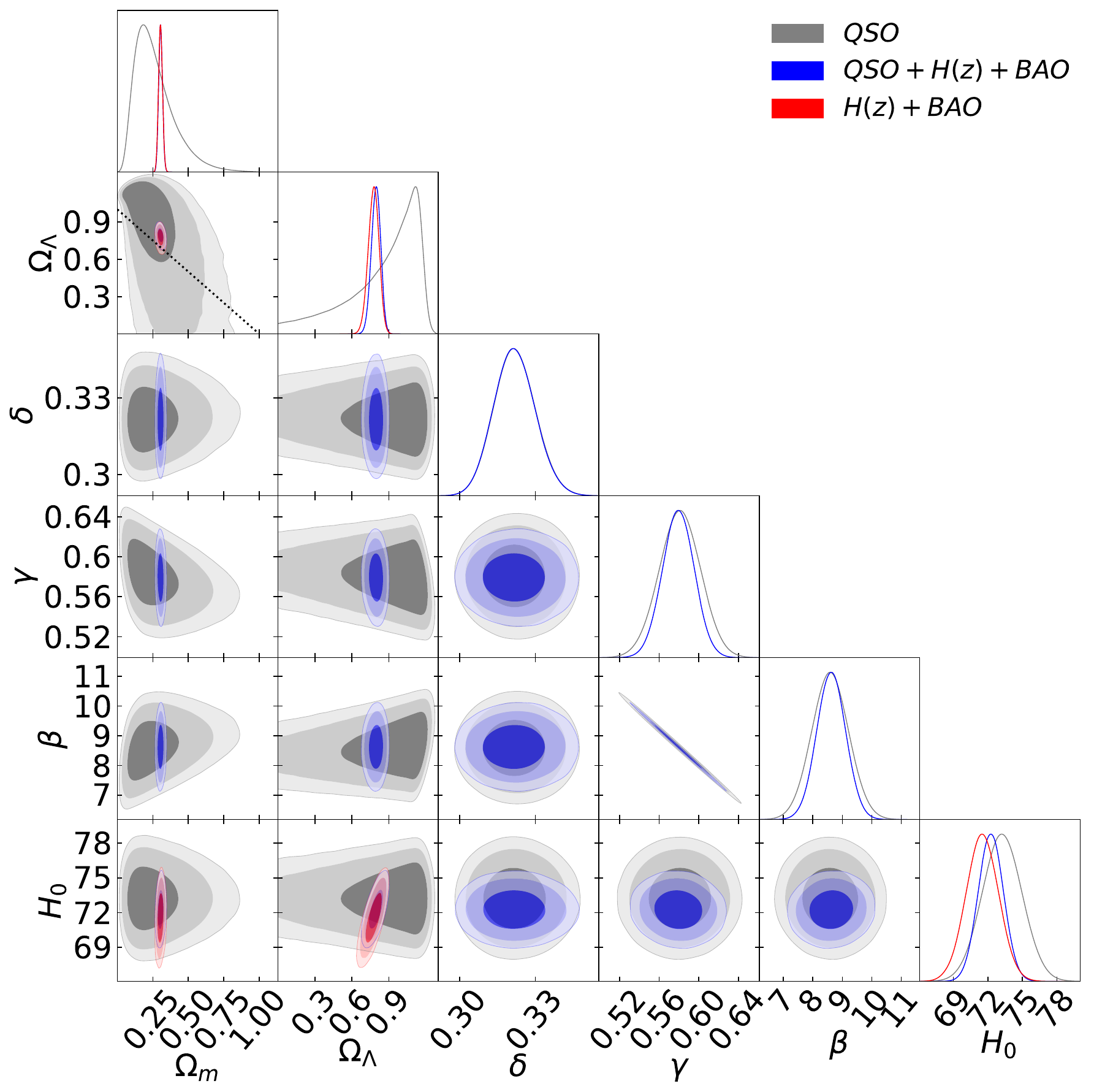}\par
    \includegraphics[width=\linewidth]{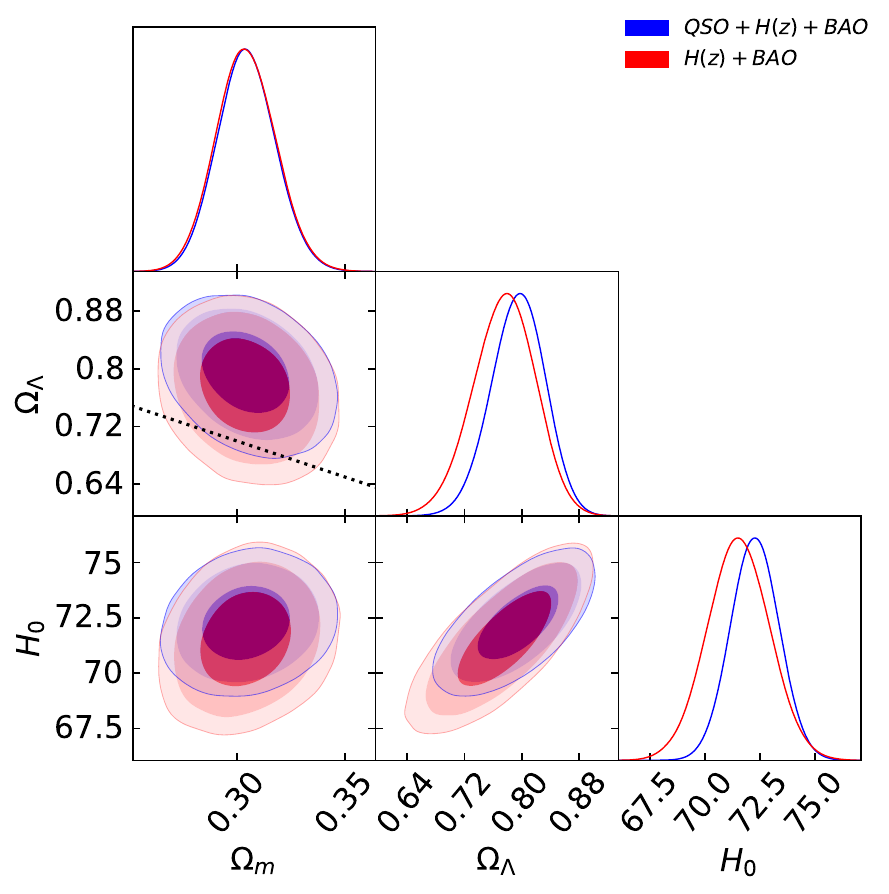}\par
\end{multicols}
\caption{Non-flat \lcdm\ model constraints from QSO (grey), $H(z)$ + BAO (red),  and QSO + $H(z)$ + BAO (blue) data. Left panel shows 1, 2, and 3$\sigma$ confidence contours and one-dimensional likelihoods for all free parameters. Right panel shows magnified plots for only cosmological parameters $\om$, $\ol$, and $H_0$, without the QSO-only constraints. These plots are for the $H_0 = 73.24 \pm 1.74$ ${\rm km}\hspace{1mm}{\rm s}^{-1}{\rm Mpc}^{-1}$ prior. The black dotted straight lines corresponds to the flat $\Lambda$CDM model.}
\label{fig:non-flat LCDM73 model with BAO, H(z) and QSO data}
\end{figure*}
\begin{figure*}
\begin{multicols}{1}
    \includegraphics[width=\linewidth]{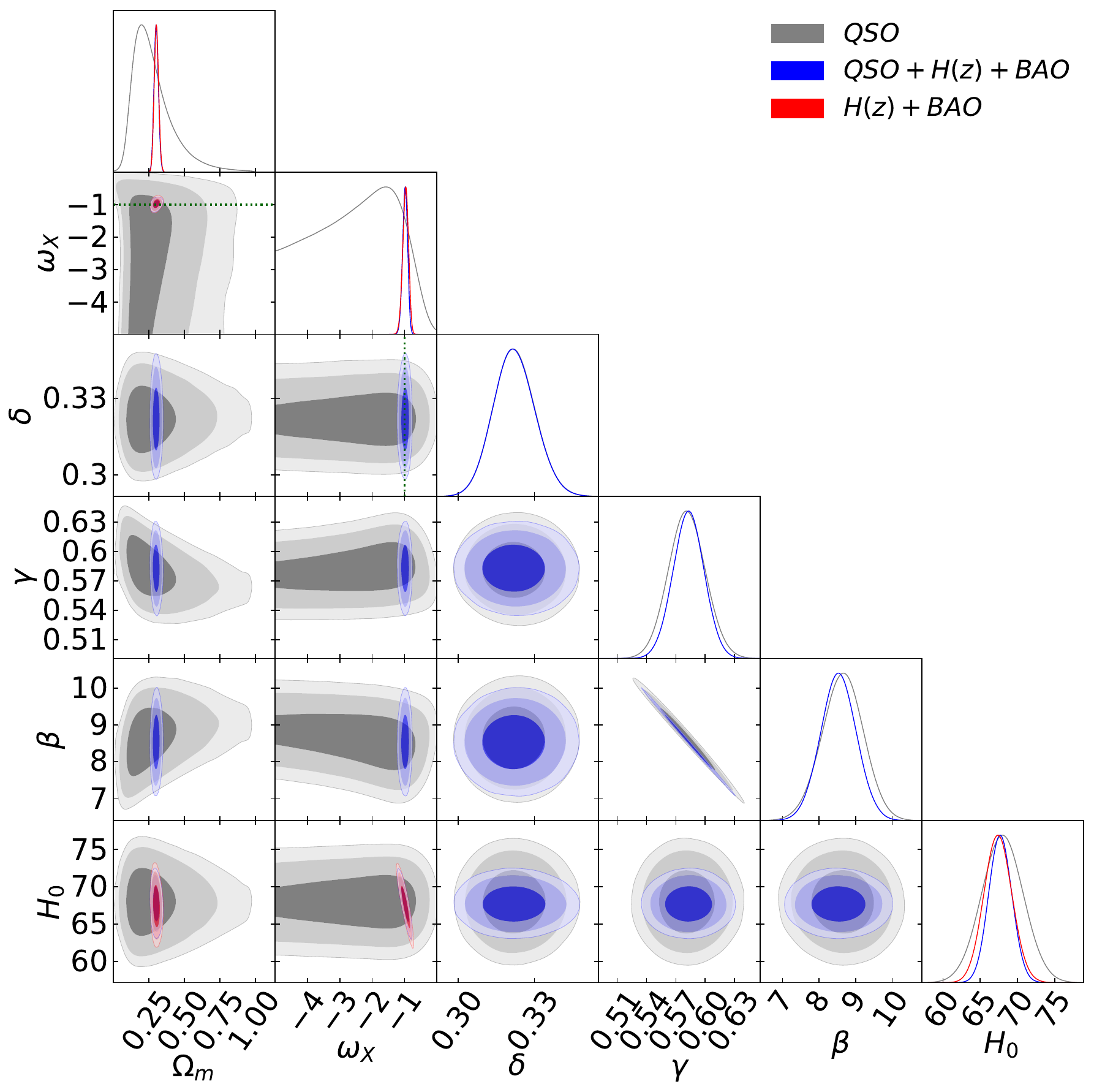}\par
    \includegraphics[width=\linewidth]{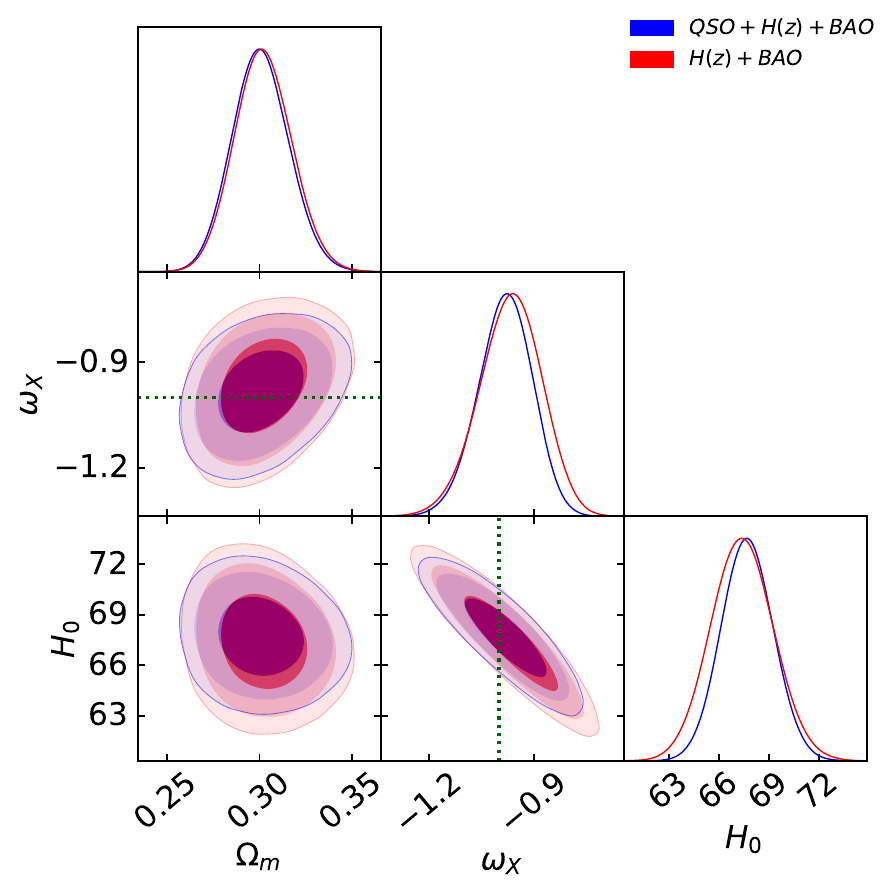}\par
\end{multicols}
\caption{Flat XCDM model constraints from QSO (grey), $H(z)$ + BAO (red),  and QSO + $H(z)$ + BAO (blue) data. Left panel shows 1, 2, and 3$\sigma$ confidence contours and one-dimensional likelihoods for all free parameters. Right panel shows magnified plots for only cosmological parameters $\om$, $\omega_X$, and $H_0$, without the QSO-only constraints. These plots are for the $H_0 = 68 \pm 2.8$ ${\rm km}\hspace{1mm}{\rm s}^{-1}{\rm Mpc}^{-1}$ prior. The green dotted straight lines represent $\omega_x$ = $-1$.}
\label{fig:Flat XCDM68 model with BAO, H(z) and QSO data}
\end{figure*}
\begin{figure*}
\begin{multicols}{1}
    \includegraphics[width=\linewidth]{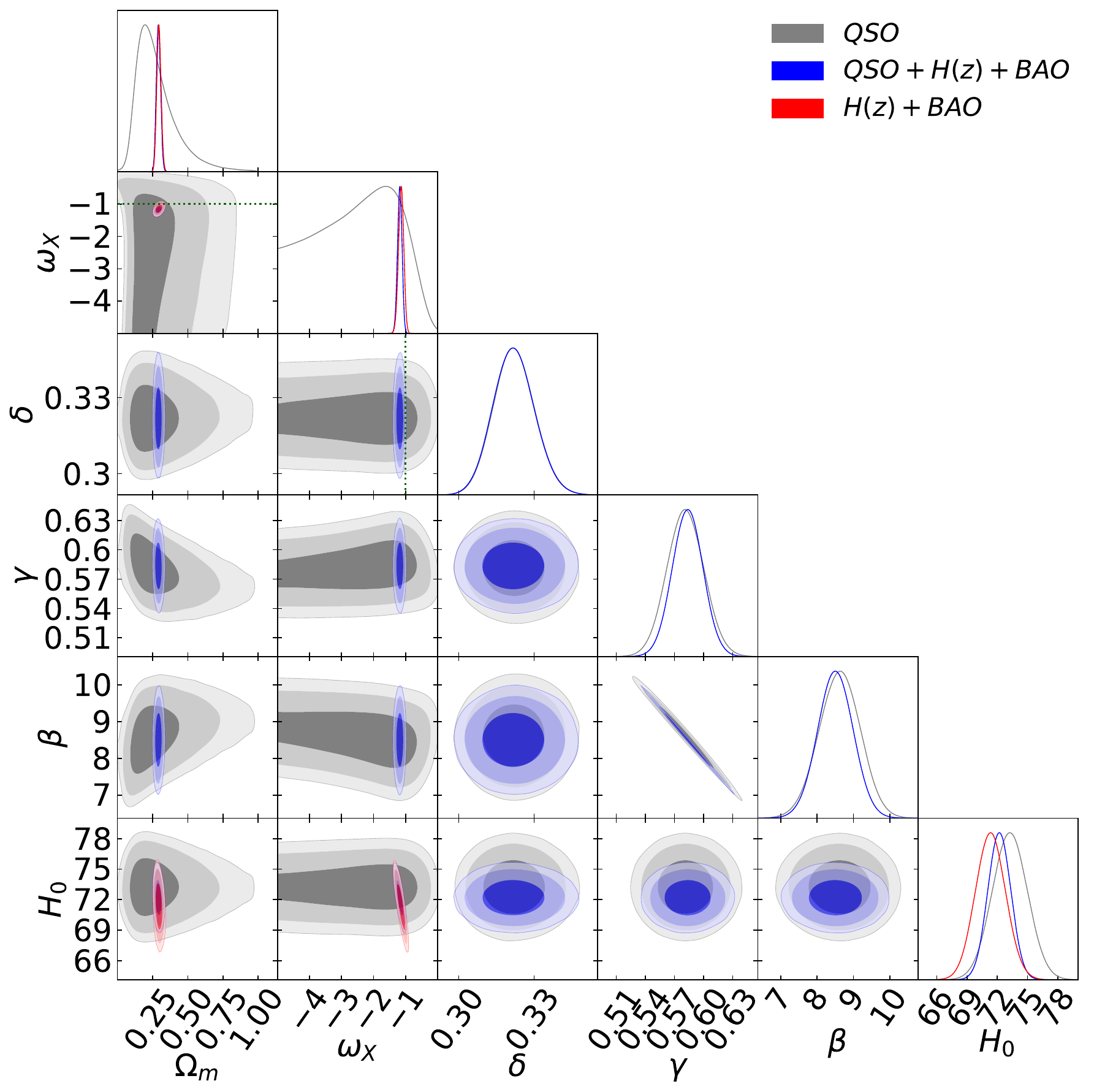}\par
    \includegraphics[width=\linewidth]{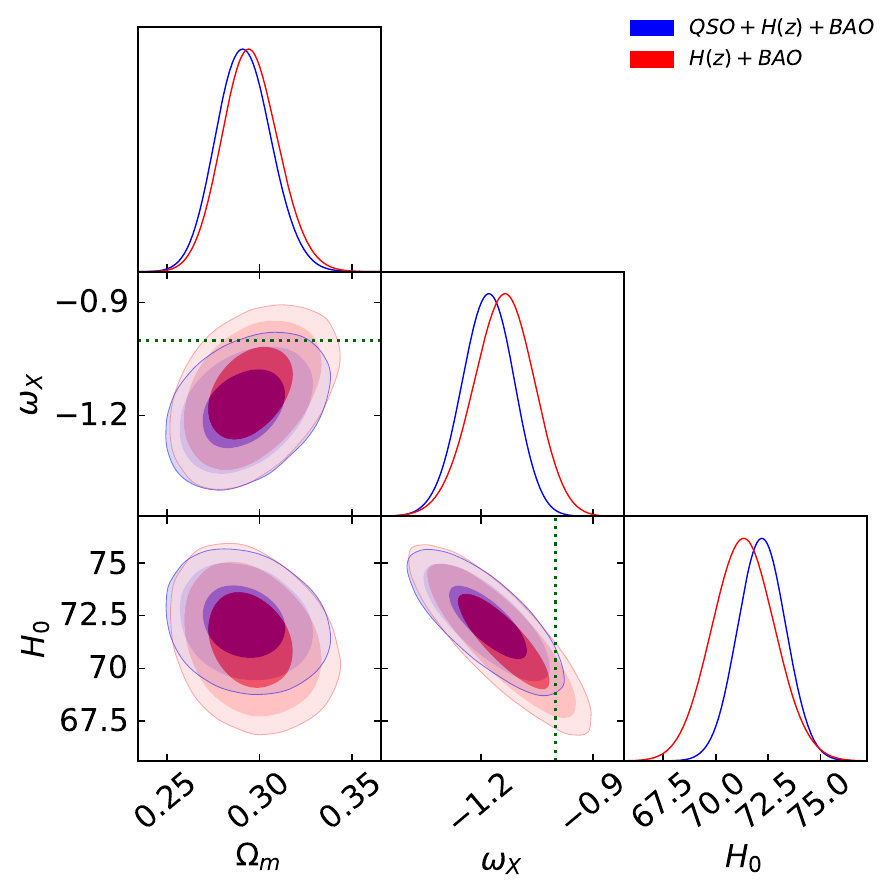}\par
\end{multicols}
\caption{Flat XCDM model constraints from QSO (grey), $H(z)$ + BAO (red),  and QSO + $H(z)$ + BAO (blue) data. Left panel shows 1, 2, and 3$\sigma$ confidence contours and one-dimensional likelihoods for all free parameters. Right panel shows magnified plots for only cosmological parameters $\om$, $\omega_X$, and $H_0$, without the QSO-only constraints. These plots are for the $H_0 = 73.24 \pm 1.74$ ${\rm km}\hspace{1mm}{\rm s}^{-1}{\rm Mpc}^{-1}$ prior. The green dotted straight lines represent $\omega_x$ = $-1$.}
\label{fig:Flat XCDM73 model with BAO, H(z) and QSO data}
\end{figure*}
\begin{figure*}
\begin{multicols}{1}
    \includegraphics[width=\linewidth]{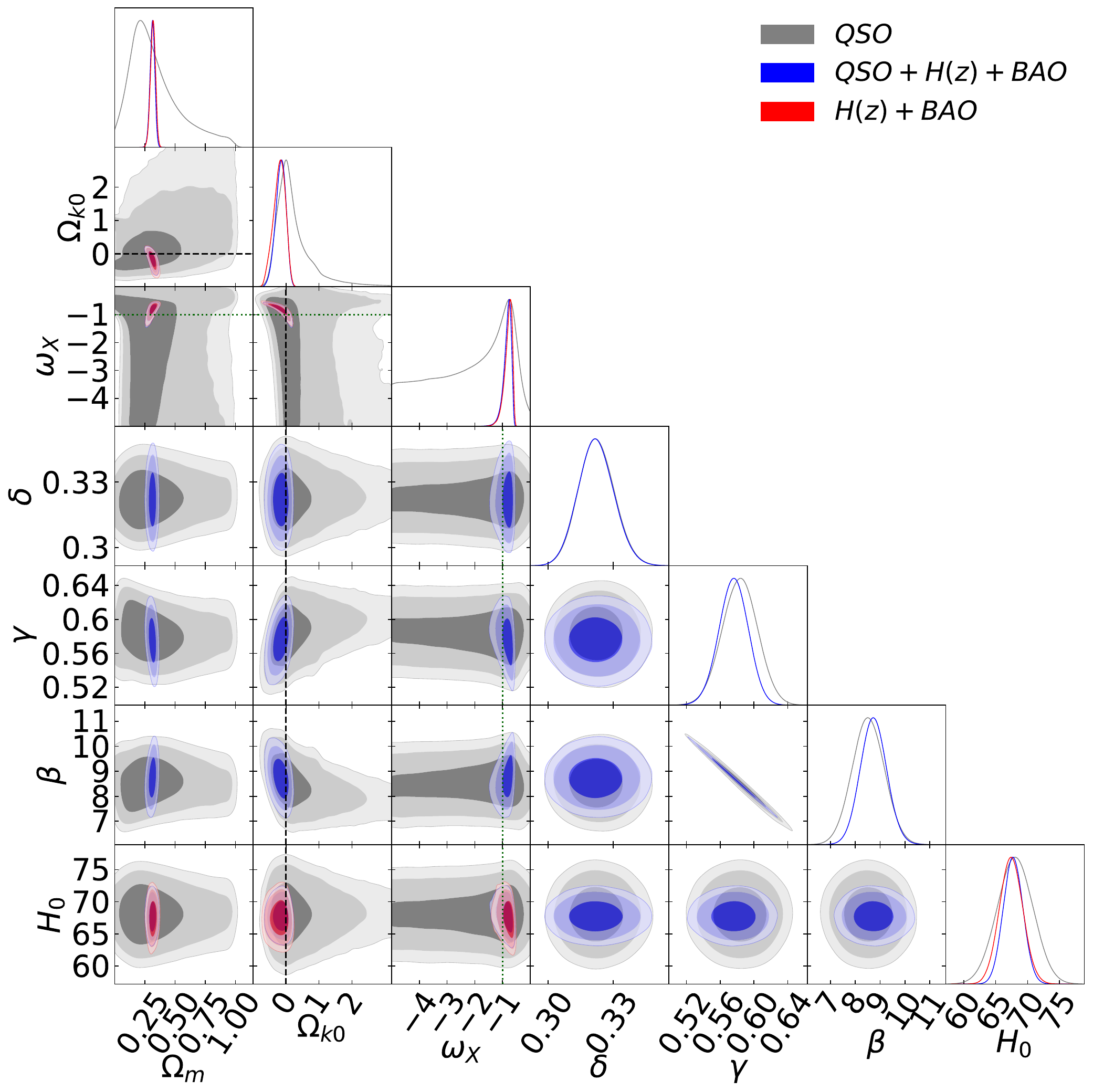}\par
    \includegraphics[width=\linewidth]{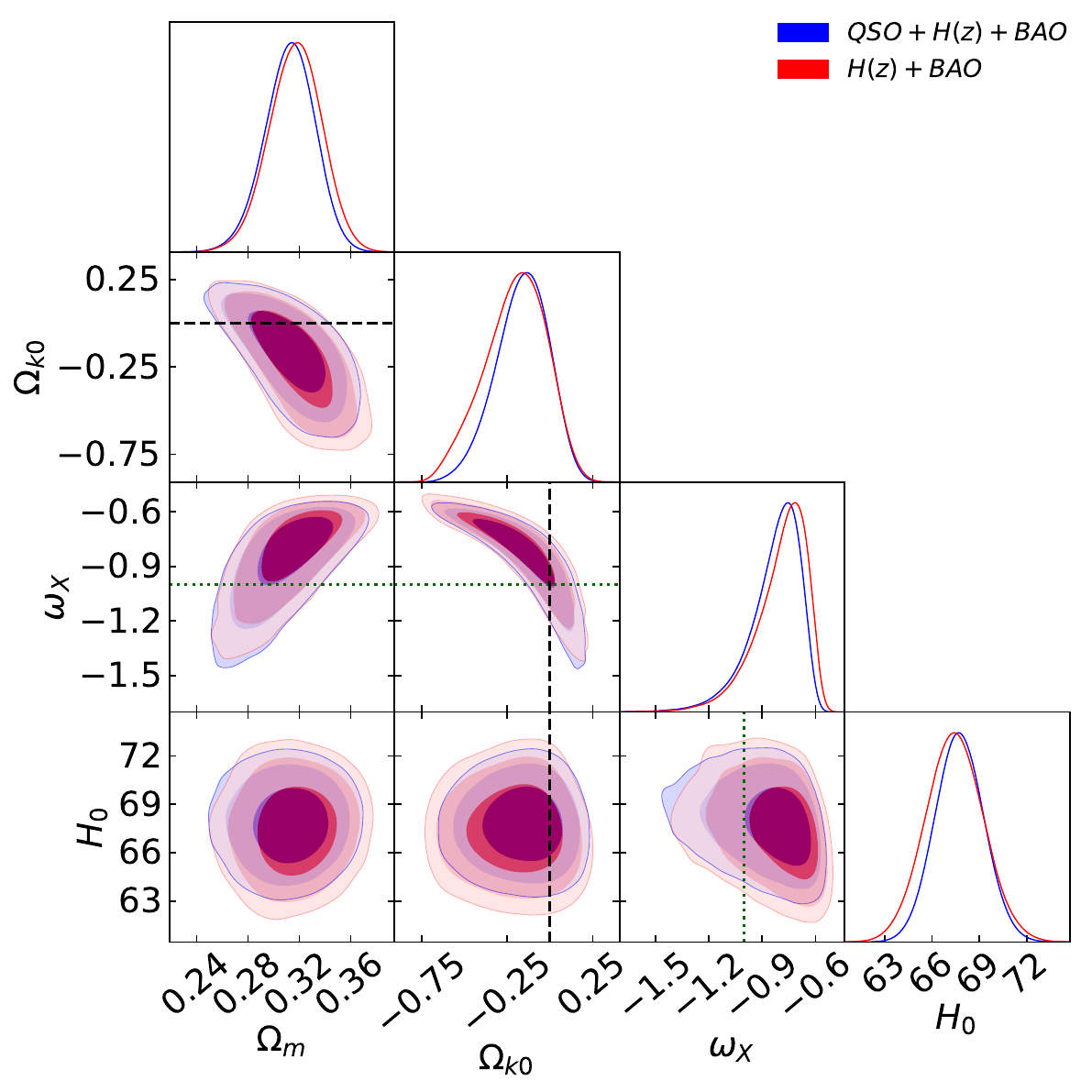}\par
\end{multicols}
\caption{Non-flat XCDM model constraints from QSO (grey), $H(z)$ + BAO (red),  and QSO + $H(z)$ + BAO (blue) data. Left panel shows 1, 2, and 3$\sigma$ confidence contours and one-dimensional likelihoods for all free parameters. Right panel shows magnified plots for only cosmological parameters $\om$, $\ok$, $\omega_X$, and $H_0$, without the QSO-only constraints. These plots are for the $H_0 = 68 \pm 2.8$ ${\rm km}\hspace{1mm}{\rm s}^{-1}{\rm Mpc}^{-1}$ prior. The black dashed straight lines and the green dotted straight lines are $\ok$ = 0 and $\omega_x$ = $-1$ lines.}
\label{fig:non-flat XCDM68 model with BAO, H(z) and QSO data}
\end{figure*}
\begin{figure*}
\begin{multicols}{1}
    \includegraphics[width=\linewidth]{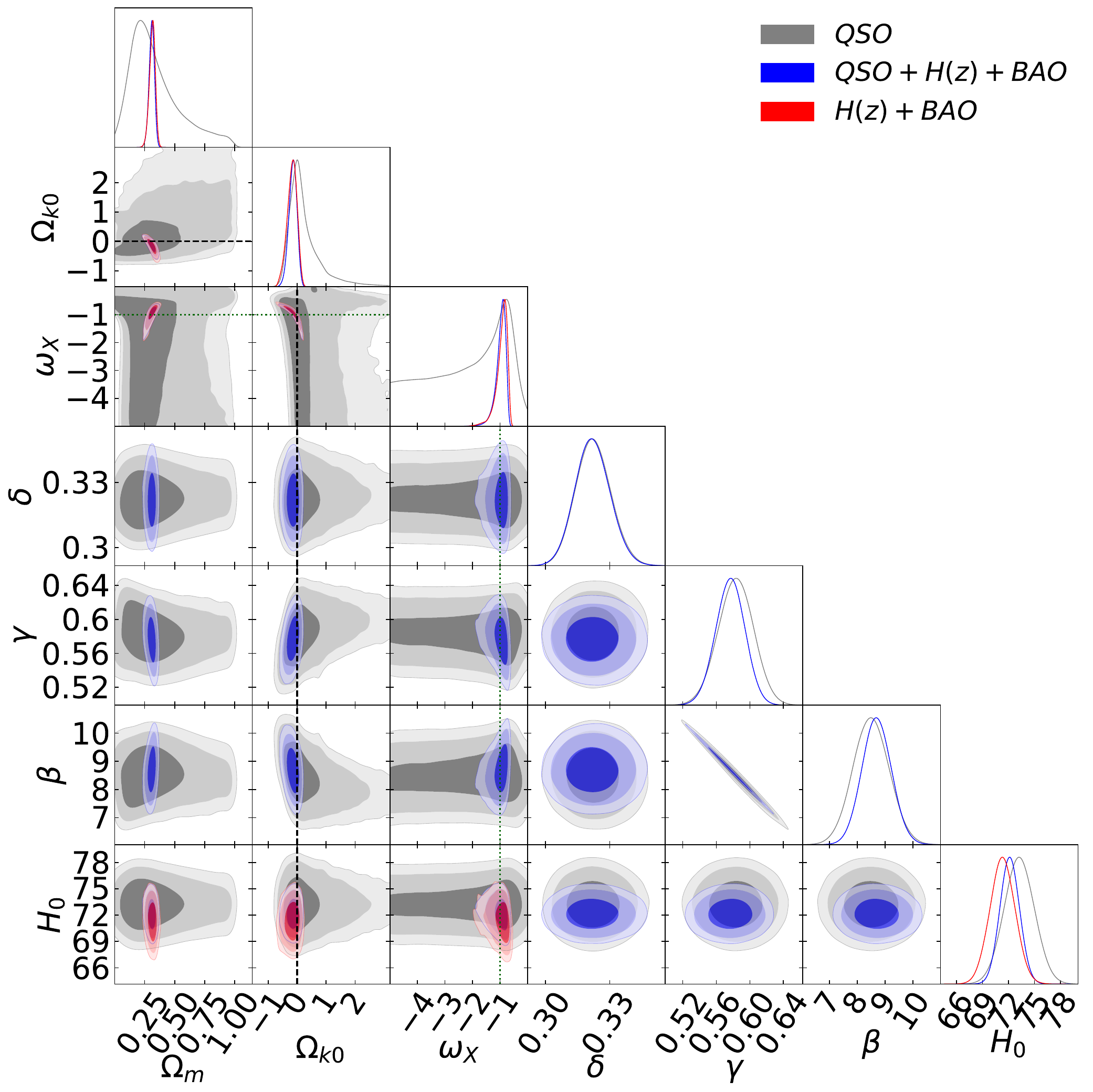}\par
    \includegraphics[width=\linewidth]{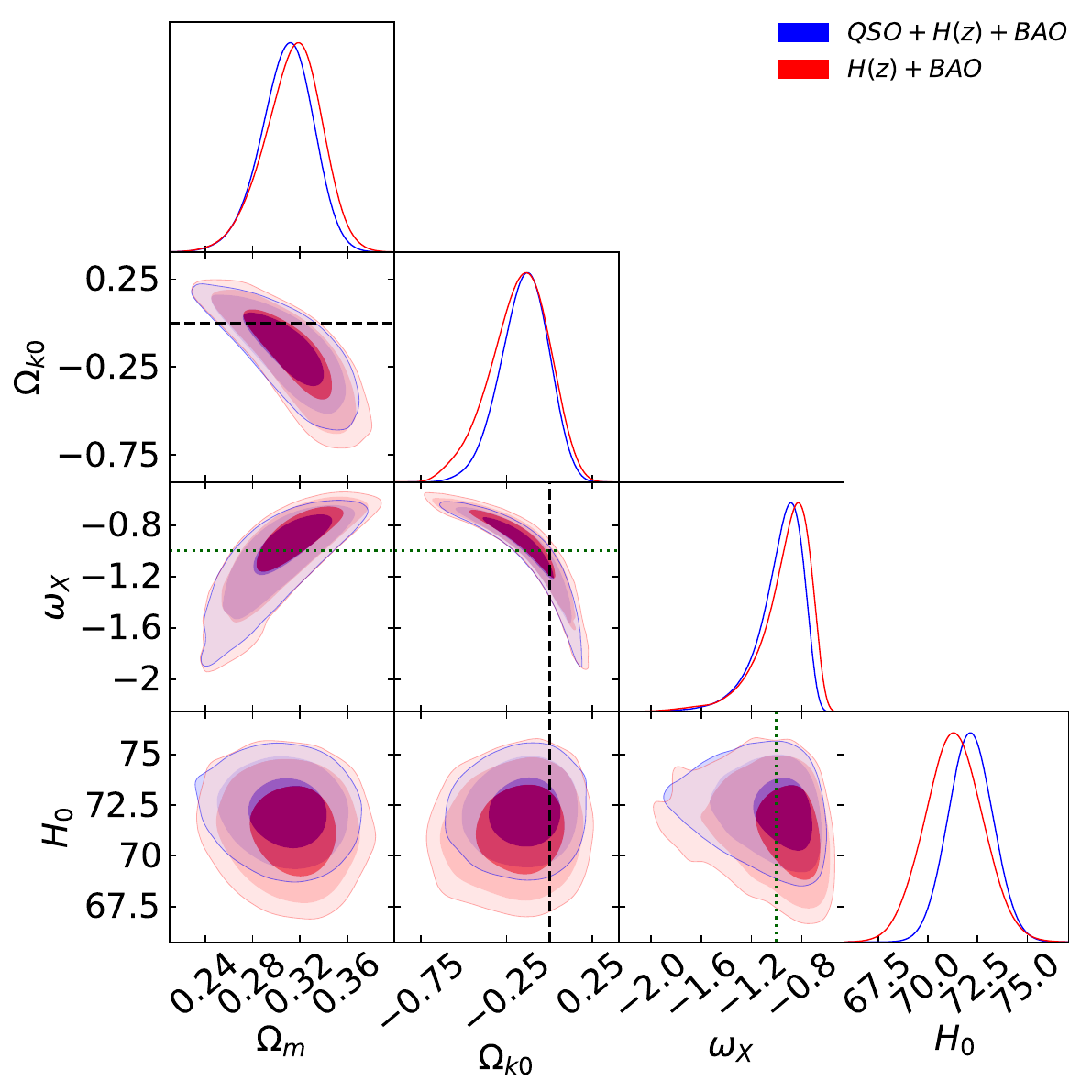}\par
\end{multicols}
\caption{Non-flat XCDM model constraints from QSO (grey), $H(z)$ + BAO (red),  and QSO + $H(z)$ + BAO (blue) data. Left pnnel shows 1, 2, and 3$\sigma$ confidence contours and one-dimensional likelihoods for all free parameters. Right panel shows magnified plots for only cosmological parameters $\om$, $\ok$, $\omega_X$, and $H_0$, without the QSO-only constraints. These plots are for the $H_0 = 73.24 \pm 1.74$ ${\rm km}\hspace{1mm}{\rm s}^{-1}{\rm Mpc}^{-1}$ prior. The black dashed straight lines and the green dotted straight lines are $\ok$ = 0 and $\omega_x$ = $-1$ lines.}
\label{fig:non-flat XCDM68 model with BAO, H(z) and QSO data}
\end{figure*}
\begin{figure*}
\begin{multicols}{1}
    \includegraphics[width=\linewidth]{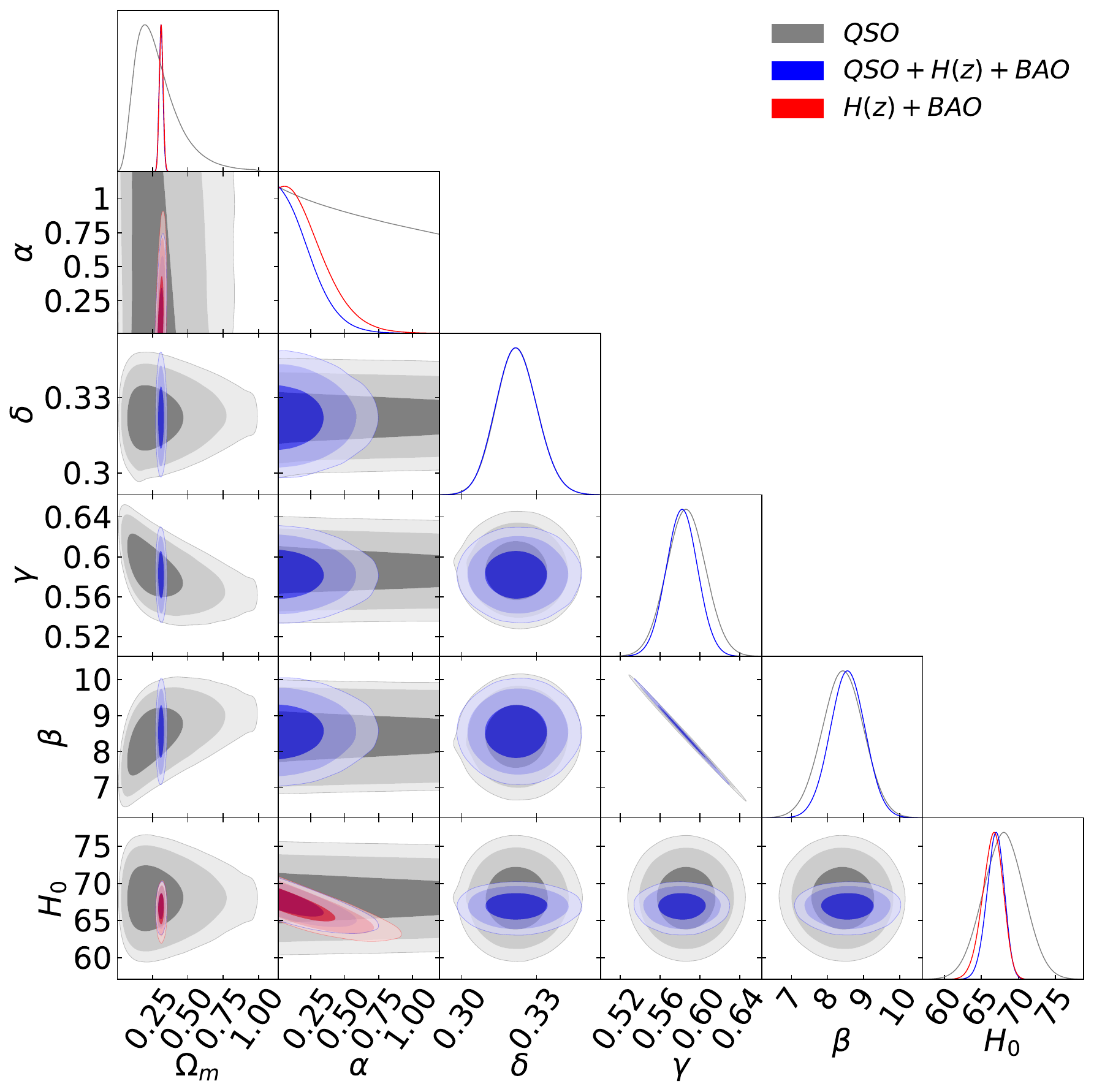}\par
    \includegraphics[width=\linewidth]{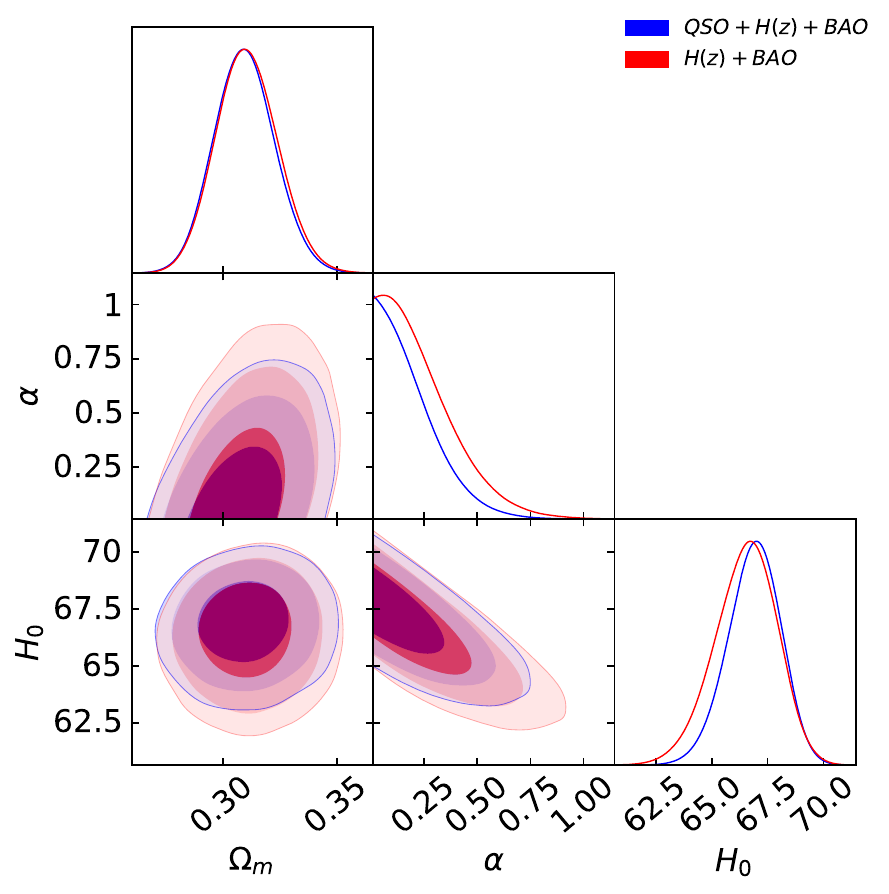}\par
\end{multicols}
\caption{Flat \pcdm\ model constraints from QSO (grey), $H(z)$ + BAO (red),  and QSO + $H(z)$ + BAO (blue) data. Left panel shows 1, 2, and 3$\sigma$ confidence contours and one-dimensional likelihoods for all free parameters. Right panel shows magnified plots for only cosmological parameters $\om$, $\alpha$, and $H_0$, without the QSO-only constraints. These plots are for the $H_0 = 68 \pm 2.8$ ${\rm km}\hspace{1mm}{\rm s}^{-1}{\rm Mpc}^{-1}$ prior.}
\label{fig:flat fphiCDM68 model with BAO, H(z) and QSO data}
\end{figure*}
\begin{figure*}
\begin{multicols}{1}
    \includegraphics[width=\linewidth]{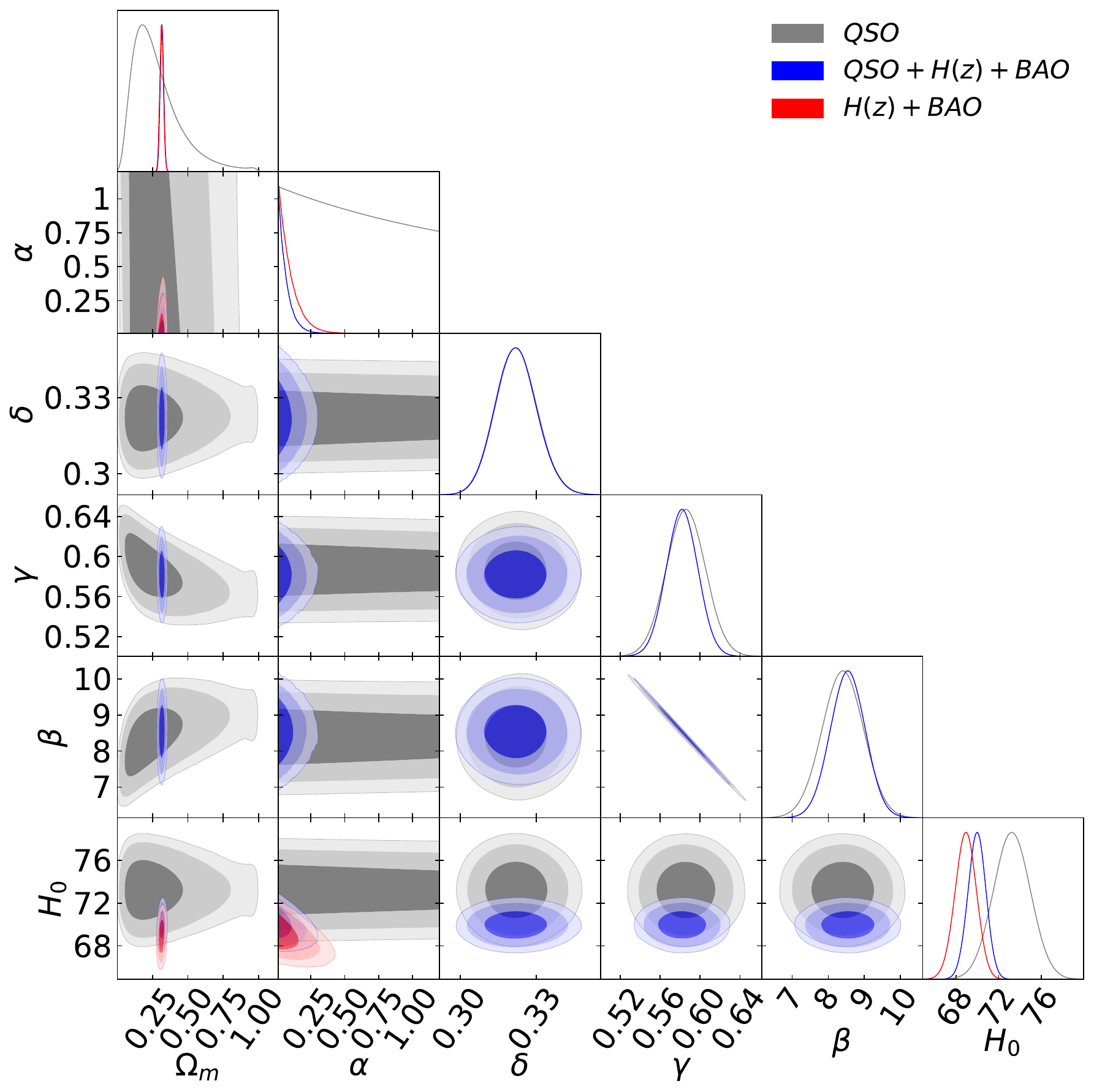}\par
    \includegraphics[width=\linewidth]{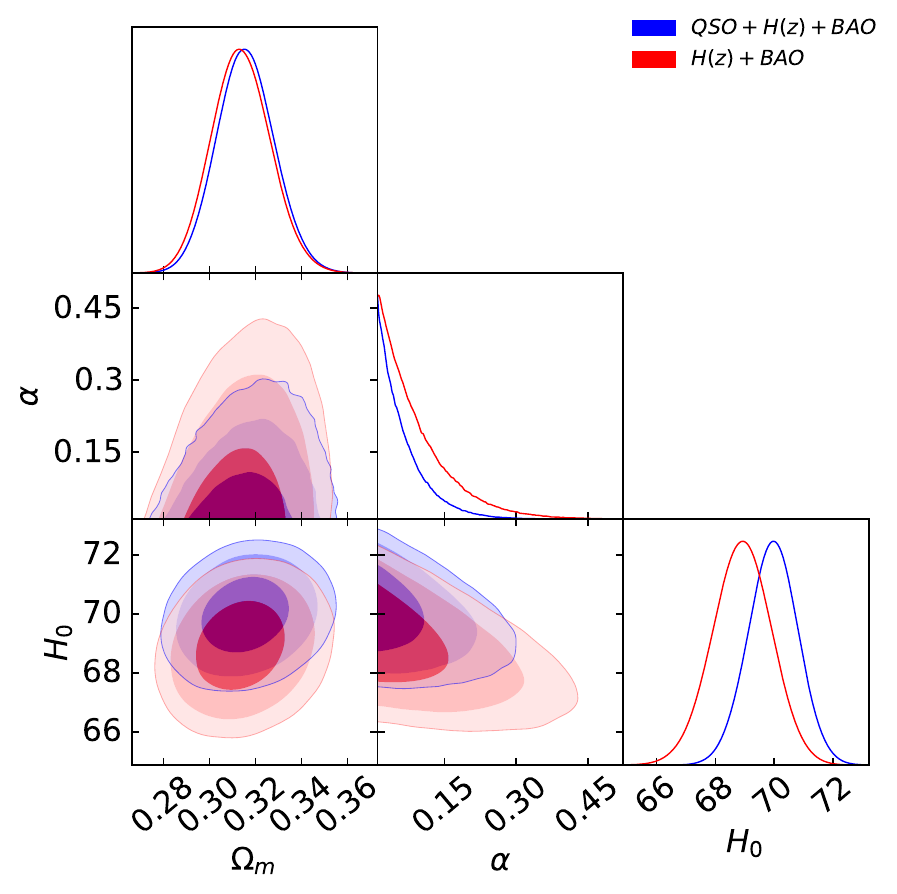}\par
\end{multicols}
\caption{Flat \pcdm\ model constraints from QSO (grey), $H (z)$ + BAO (red),  and QSO + $H(z)$ + BAO (blue) data. Left panel shows 1, 2, and 3$\sigma$ confidence contours and one-dimensional likelihoods for all free parameters. Right panel shows magnified plots for only cosmological parameters $\om$, $\alpha$, and $H_0$, without the QSO-only constraints. These plots are for the $H_0 = 73.24 \pm 1.74$ ${\rm km}\hspace{1mm}{\rm s}^{-1}{\rm Mpc}^{-1}$ prior.}
\label{fig:flat fphiCDM73 model with BAO, H(z) and QSO data}
\end{figure*}
\begin{figure*}
\begin{multicols}{1}
    \includegraphics[width=\linewidth]{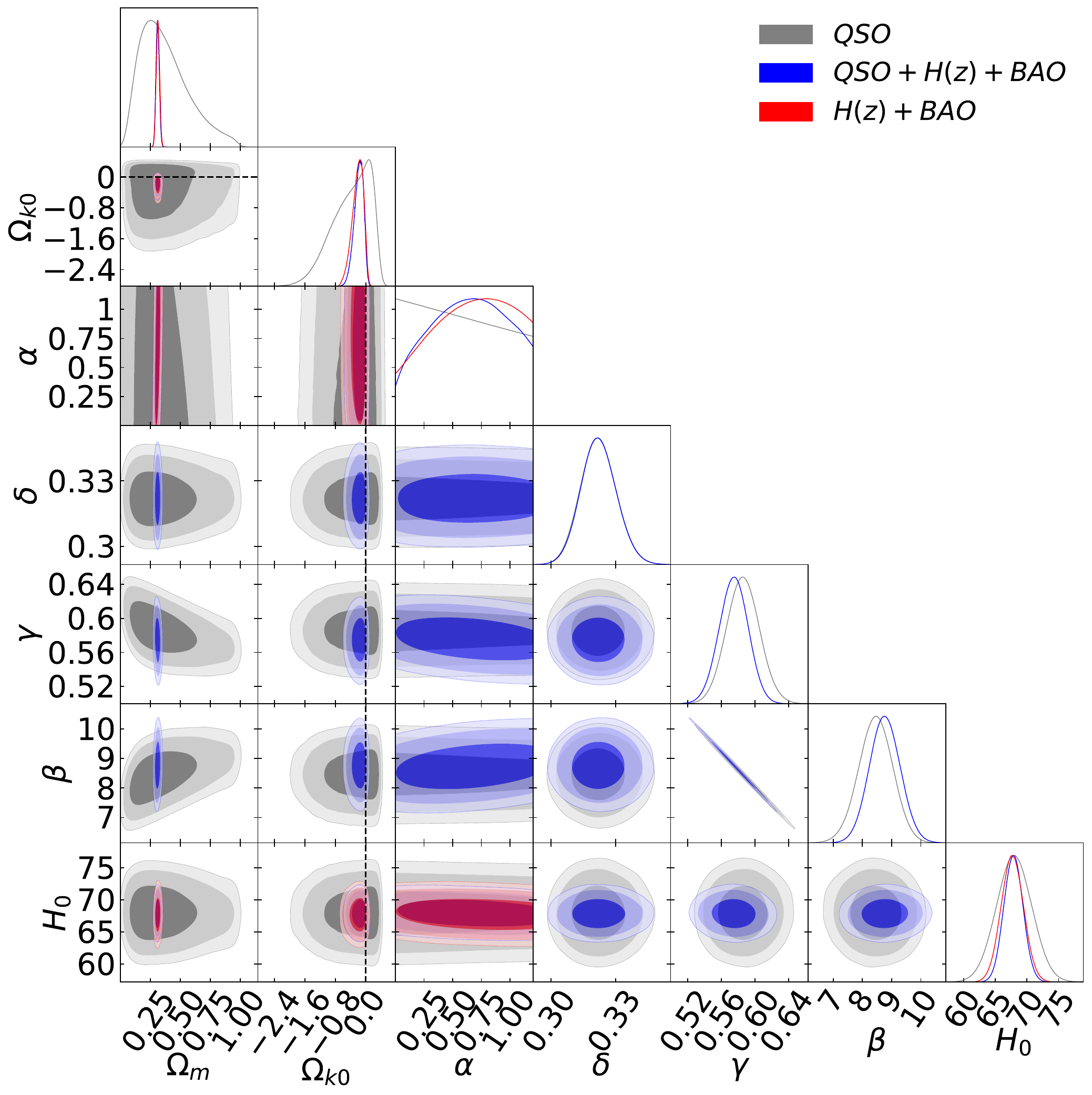}\par
    \includegraphics[width=\linewidth]{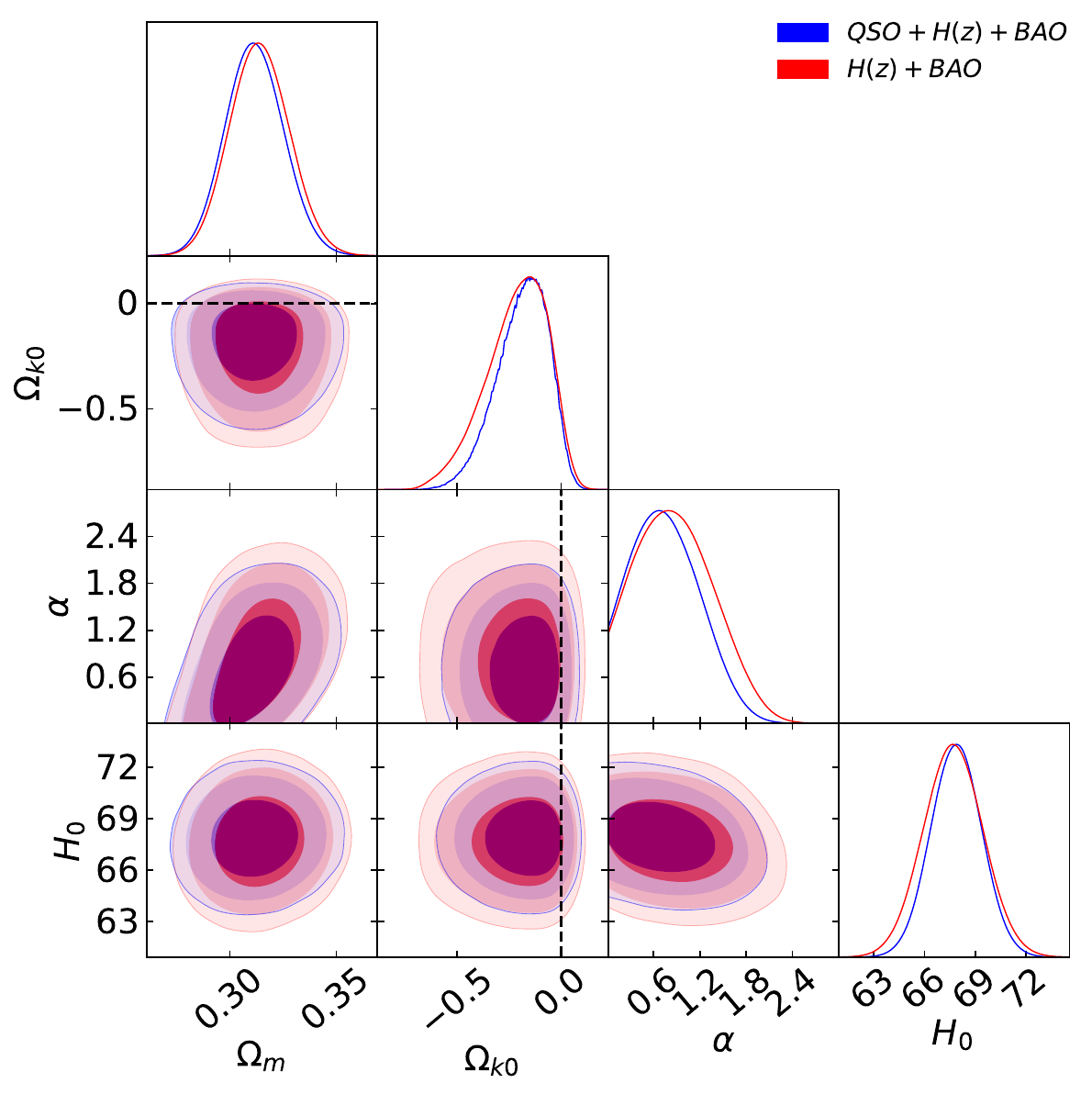}\par
\end{multicols}
\caption{Non-flat \pcdm\ model constraints from QSO (grey), $H(z)$ + BAO (red),  and QSO + $H(z)$ + BAO (blue) data. Left panel shows 1, 2, and 3$\sigma$ confidence contours and one-dimensional likelihoods for all free parameters. Right panel shows magnified plots for only cosmological parameters  $\om$, $\ok$, $\alpha$, and $H_0$, without the QSO-only constraints. These plots are for the $H_0 = 68 \pm 2.8$ ${\rm km}\hspace{1mm}{\rm s}^{-1}{\rm Mpc}^{-1}$ prior. The black dashed straight lines are $\ok$ = 0 lines.}
\label{fig: nfphiCDM73 model with BAO, H(z) and QSO data}
\end{figure*}
\begin{figure*}
\begin{multicols}{1}
    \includegraphics[width=\linewidth]{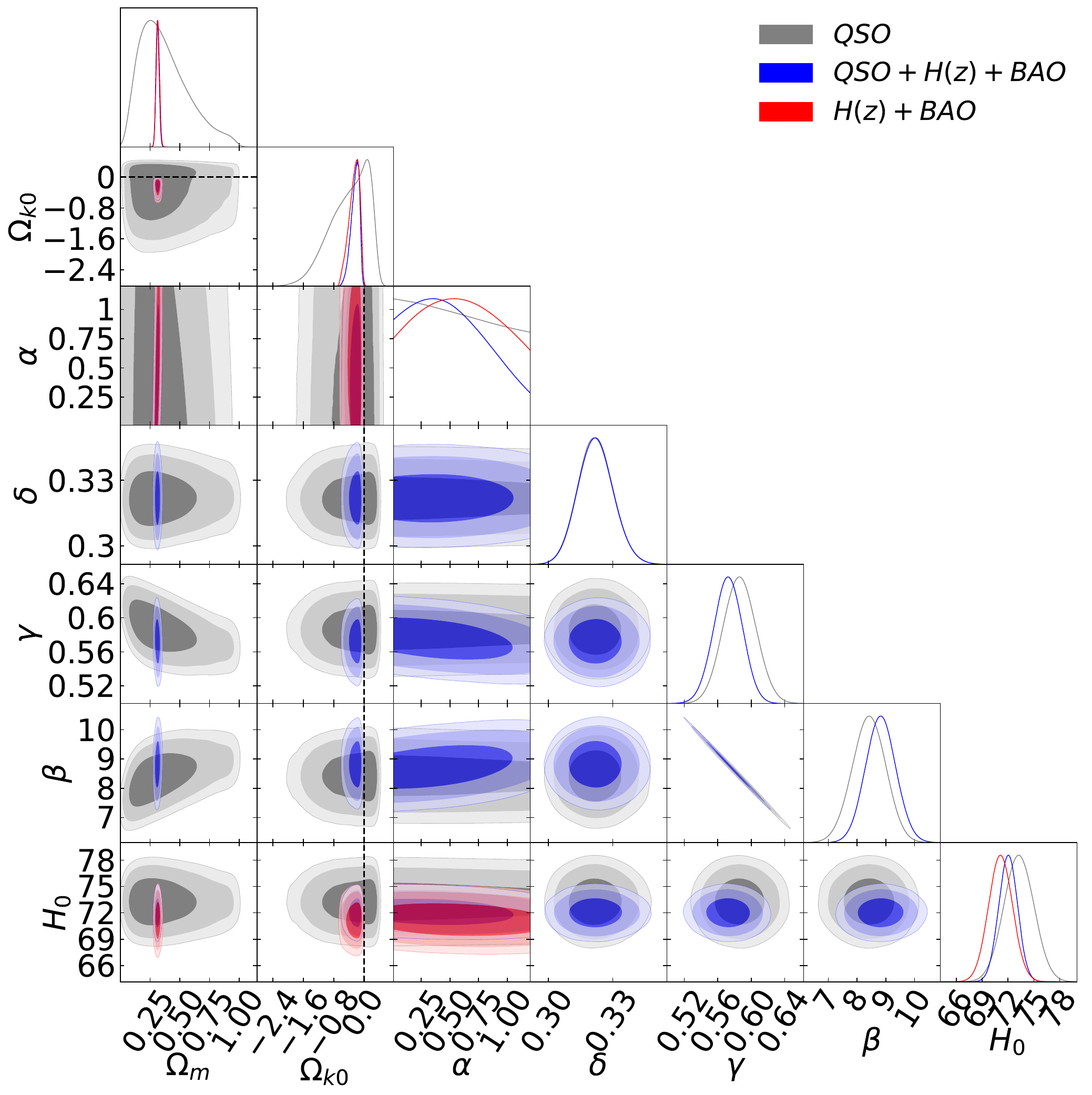}\par
    \includegraphics[width=\linewidth]{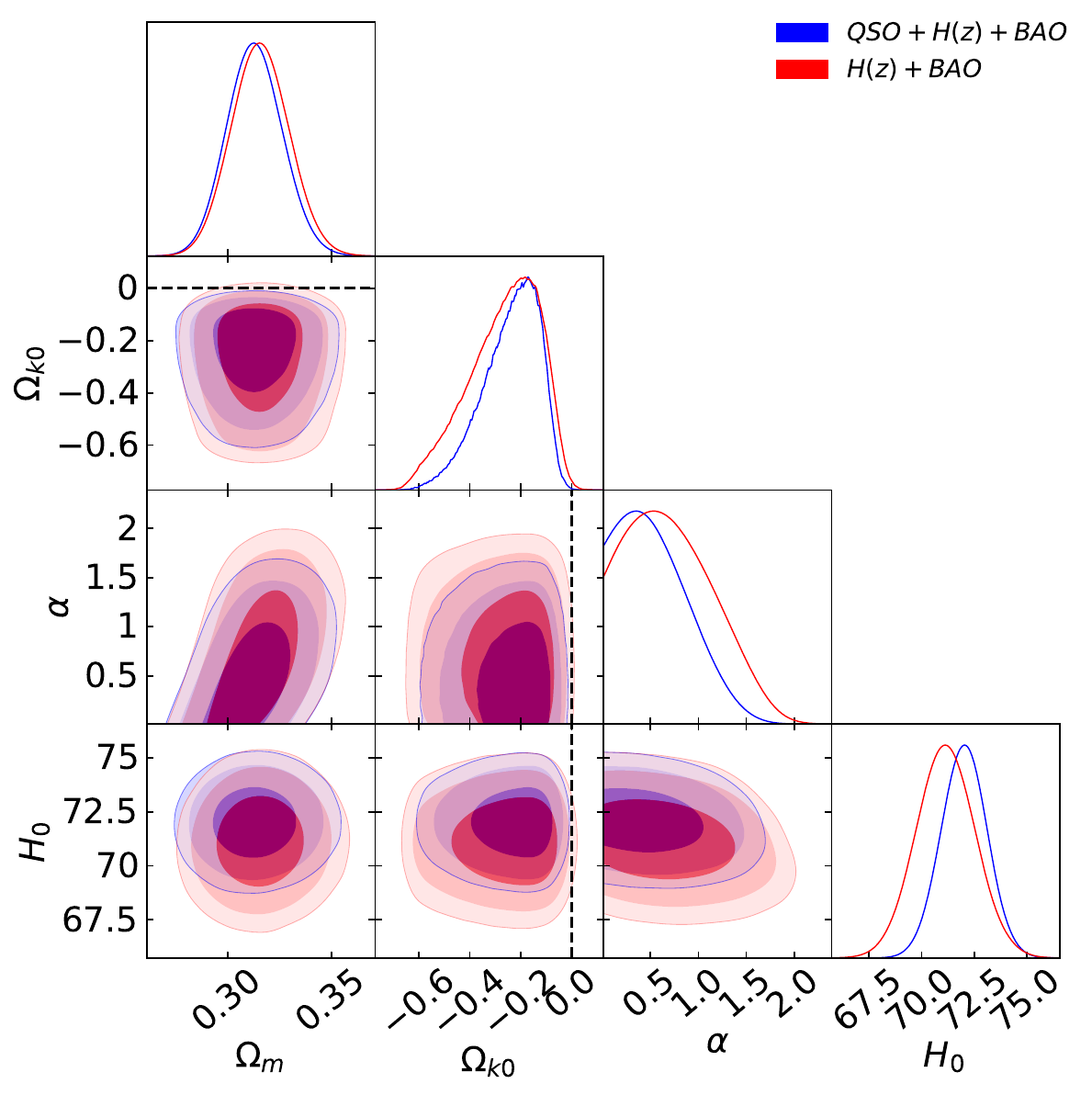}\par
\end{multicols}
\caption{Non-Flat \pcdm\ model constraints from QSO (grey), $H(z)$ + BAO (red),  and QSO + $H(z)$ + BAO (blue) data. Left panel shows 1, 2, and 3$\sigma$ confidence contours and one-dimensional likelihoods for all free parameters. Right panel shows magnified plots for only cosmological parameters $\om$, $\ok$, $\alpha$, and $H_0$, without the QSO-only constraints.These plots are the for $H_0 = 73.24 \pm 1.74$ ${\rm km}\hspace{1mm}{\rm s}^{-1}{\rm Mpc}^{-1}$ prior. The black dashed straight lines are $\ok$ = 0 lines.}
\label{fig: nfphiCDM73 model with BAO, H(z) and QSO data}
\end{figure*}
%%%%%%%%%%%%%%%%%%%%%%%%%%%%%%%%%%%%%%%%%%%%%%%%%%

%%%%%%%%%%%%%%%%% APPENDICES %%%%%%%%%%%%%%%%%%%%%
\begin{comment}
\appendix

\newpage~

\newpage~

\newpage~

\newpage~

\newpage~

\newpage~

\newpage~

\newpage~

\newpage~

\newpage~

\end{comment}

%%%%%%%%%%%

% Don't change these lines
\bsp	% typesetting comment
\label{lastpage}
\end{document}